\documentclass{iopart}

\newcommand{\one}{\mbox{$1 \hspace{-1.0mm}  {\bf l}$}}
\newcommand{\ket}[1]{\left|{#1}\right\rangle}

\newenvironment{multline*}{\begin{eqnarray}}{\end{eqnarray}}

\usepackage{dcolumn} 
\usepackage{amssymb,graphicx,amstext}

\begin{document}
\bibliographystyle{iop}


\title{Entanglement and its dynamics in open, dissipative systems}

\author{L. Hartmann$^{1}$, W. D{\"u}r$^{1,2}$ and H.-J. Briegel$^{1,2}$}

\address{$^1$ Institut f{\"u}r Theoretische Physik, Universit{\"a}t Innsbruck,
Technikerstra{\ss}e 25, A-6020 Innsbruck, Austria\\
$^2$ Institut f\"ur Quantenoptik und Quanteninformation der \"Osterreichischen Akademie der Wissenschaften, Innsbruck, Austria.}
\date{\today}

\begin{abstract}
Quantum mechanical entanglement can exist in noisy open quantum systems at high temperature. A simple mechanism, where system particles are randomly reset to some standard initial state, can counteract the deteriorating effect of decoherence, resulting in an entangled steady state far from thermodynamical equilibrium. We present models for both gas-type systems and for strongly coupled systems. We point out in which way the entanglement resulting from such a reset mechanism is different from the entanglement that one can find in thermal states. We develop master equations to describe the system and its interaction with an environment, study toy models with two particles (qubits), where the master equation can often be solved analytically, and finally examine larger systems with possibly fluctuating particle numbers.
We find that in gas-type systems, the reset mechanism can produce an entangled steady state for an arbitrary temperature of the environment, while this is not true in strongly coupled systems. But even then, the temperature range where one can find entangled steady states is typically much higher with the reset mechanism.
\end{abstract}

%
%
%

\section{Introduction}

In quantum information theory entanglement between parts of a system has been identified as the key resource that can possibly make quantum information processing more powerful than classical information processing. Entanglement can also be a resource for long-distance quantum communication or distributed quantum computation, and it is at the heart of some quantum communication protocols. But entanglement is fragile under the influence of environment induced decoherence. All engineering hence thrives to better control and manipulate the quantum information stored in the system while keeping the detrimental effects of decoherence low.

In nature, on the other hand, we mostly find less controllable systems, especially if the system size becomes macroscopic as in gases, fluids, solids or even biological systems. Since these systems are usually open, noisy systems at possibly high temperatures one expects that environment induced decoherence will erase all entanglement between system degrees of freedom. This reasoning is true except for three cases.

First, the environment and its coupling to the system could be special in a way that it creates rather than destroys entanglement. However, it is unlikely to find such an environment in nature where usually thermalization dominates, and we will only briefly touch upon the subject of such environments in this paper.

Second, if the system has an entangled ground state, as many solid state systems do, its thermal state will be entangled in a certain temperature range above zero by a continuity argument. Coupling to a heat bath drives a system into its thermal state. But there is a temperature threshold for the bath above which the thermal state will be unentangled.

Third, the system might have a built-in entropy drain, meaning that the correlations with the environment are, by one way or another, erased such that the system can re-build entanglement through its quantum mechanical interactions. This entropy drain may even be local to exclude the trivial cases where entanglement is simply ``pumped" into the system, e.g., by injecting fresh, entangled Bell pairs.

In~\cite{hartmann_reset} we proposed such a local entropy drain in form of a reset mechanism, where system particles are randomly replaced by particles in some standard, mixed state of sufficiently low entropy. Note that such a mechanism cannot create entanglement, on the contrary, it erases any entanglement that might still be present between the particle that is reset and the rest of the system. Only the interplay with the system Hamiltonian can lead to entanglement in a steady state that is possibly far from thermodynamic equilibrium. This reset mechanism was studied for a toy model with two qubits, where analytic solutions could be obtained. Also a multipartite scenario for a (simplified) gas model was discussed, and further generalizations were suggested. By gas-type systems we mean systems in which the decoherence processes act locally on the system particles, by strongly coupled systems we mean those where the decoherence processes act globally. 
To be more precise, local decoherence processes are those, which induce transitions between the eigenstates of the local, free Hamiltonian alone, while global decoherence processes induce transitions between eigenstates of the total Hamiltonian.

In this paper we review the key idea of a reset mechanism but provide more in-depth material than in~\cite{hartmann_reset}. We elaborate on the generalizations suggested in~\cite{hartmann_reset}, namely on the influence of local entropy drains on the dynamics of entanglement and on the steady-state entanglement in gas-type systems as well as in strongly coupled systems.

We prove that the master equation describing the evolution of the system coupled to a heat bath and subject to a reset mechanism is of Lindblad form and hence generates a completely positive, i.e., physical map. 
We analytically solve the master equation for small systems of two spins with special interaction Hamiltonians, which enables us to illustrate the main features of the reset mechanism. In particular we show the following.

\begin{enumerate}
\item Steady-state entanglement in systems with reset mechanism is different from the entanglement in thermal states.
\item In strongly coupled systems with constant coupling steady-state entanglement with reset can exist for higher temperatures than the entanglement in the thermal state, which is the steady state without reset.
\item In gas-type systems steady-state entanglement with reset can exist even for arbitrary temperatures.
\end{enumerate}

These features are not due to the specially chosen interaction Hamiltonians and decoherence processes. We demonstrate that the above properties are almost independent of both. One can also relax the conditions on the reset states and take mixed states with sufficiently low entropy instead of pure states. Finally, a generalization to larger system sizes, possibly even with fluctuating particle numbers, still leads to similar results. Hence, the reset mechanism is at the same time simple and generic.

We remark that in cavity QED an incoherent generation of entanglement has been proposed, which bears resemblance to our work~\cite{plenio_entangled_light}. There, an atom couples to two leaky optical cavities and is driven by a white noise field. This incoherent driving can, when the atom is finally traced out, result in entanglement between the cavity modes. Entanglement is generated for intermediate cavity damping rates and intensities of the noise field, an effect labeled  ``stochastic resonance'' in~\cite{plenio_entangled_light,huelga_stoch_resonance}. We believe that this effect is more correctly interpreted as an example for a reset mechanism. In a subsequent work~\cite{yi_entangled_atoms} strongly related to~\cite{plenio_entangled_light}, one single cavity entangles two atoms, giving yet another example for a reset mechanism even closer to the setups of this paper.

The reset mechanism is certainly not a preferred way to actively protect entanglement and mostly cannot even be compared to such strategies, but, because of its simplicity and generality, there is hope that such a mechanism may ultimately be identified in natural processes leading to an increased understanding whether entanglement can play a role in systems at high temperatures. 

The paper is organized as follows. We first concentrate on simple models with only two particles (qubits). In section~\ref{gasmodel} we motivate the description by a master equation, explain in which cases the model is valid, and study several specific Hamiltonians and noise channels analytically and others numerically. We also compare entangled steady states resulting from a reset mechanism to entangled steady states resulting from special choices of interaction Hamiltonian and decoherence process. We show in section~\ref{stronglycoupledmodel} that we can find the same features in strongly coupled systems, and we give the conditions to be met by the reset mechanism such that entangled steady states can exist. Then, in section~\ref{multipartite}, we extend the model to include more qubits and discuss the meaning of different kinds of entanglement that we use. Finally we give a summary of the results in section~\ref{summary}.

\section{Gas-type systems}\label{gasmodel}

In this chapter we discuss a toy model with only two particles, which we take as spin-$1/2$ systems or qubits for simplicity. The toy model shows all the features that we will later find in larger systems and it has the advantage that we can show many results analytically leading to an increased understanding of the involved processes. We will formulate the equations for an arbitrary number $N$ of qubits, so that we can refer to them later in section~\ref{multipartite}.

\subsection{Master equation for gas-type systems}\label{MEwithoutreset}

In a gas particles are weakly coupled in the sense that most of the time they do not considerably interact with each other unless they collide. In the meantime they only feel their local, free Hamiltonian and are subject to individual, local decoherence processes, e.g. through interactions with thermal photons (radiative damping). If we pick a subset consisting of $N$ gas particles and consider these as the system, collisions with the remaining gas particles are another source of decoherence (non-radiative damping or dephasing).
In a master equation that models this gas-type system we replace the original, time-dependent collision Hamiltonian by an averaged, time-independent interaction Hamiltonian. Since the interaction Hamiltonian does not modify the energy landscape in this model, the local, radiative decoherence processes tend to drive the system to the thermal state of the free Hamiltonian, for which we choose the form
\begin{equation}\label{atomicHamiltonian}
 H_{\rm{free}}=\omega/2\sum_{i=1}^N\sigma_z^{(i)}\mbox{.}
\end{equation}
We leave the interaction Hamiltonian $H$ unspecified for the moment. For two qubits, we will often use the Ising Hamiltonian
\begin{equation}\label{Isinghamiltonian}
H_{\rm{Ising}}=g\sigma_z^{(1)}\sigma_z^{(2)}
\end{equation}
for analytic discussions, whereas more complicated Hamiltonians will be treated numerically.
We write the total Hamiltonian as $H_{total}=H+H_{\rm{free}}$ such that the master equation is
\begin{equation}\label{gasME}
\dot{\rho}=-i[H_{\rm{total}},\rho]+{\cal L}_{\rm{noise}}\rho,
\end{equation}
where ${\cal L}_{\rm{noise}}$ is a Liouville operator representing the noise channels. We describe the noise channels by the Lindblad operator~\cite{briegel_qo_master_equations,hein_multipartite_entanglement}
\begin{eqnarray}\label{localnoise}
{\cal L}_{\rm{noise}}\rho=\sum_{i=1}^N -\frac{B}{2}(1-s)[\sigma_+^{(i)}\sigma_-^{(i)}\rho+\rho\sigma_+^{(i)}\sigma_-^{(i)}-2\sigma_-^{(i)}\rho\sigma_+^{(i)}]\nonumber\\
-\frac{B}{2} s[\sigma_-^{(i)}\sigma_+^{(i)}\rho+\rho\sigma_-^{(i)}\sigma_+^{(i)}-2\sigma_+^{(i)}\rho\sigma_-^{(i)}]-\frac{2C-B}{4}[\rho-\sigma_z^{(i)}\rho\sigma_z^{(i)}]
\end{eqnarray}
where $\sigma_\pm=(\sigma_x \pm i\sigma_y)/2$ and the $\sigma$s are Pauli operators. Parameters $B$ and $C$ give the decay rate of inversion $\langle\frac{1+\sigma_z}{2}\rangle$ and polarization $\langle\sigma_\pm\rangle$ under the action of ${\cal L}_{\rm{noise}}$, and $s=\lim_{t\rightarrow\infty}\langle(1+\sigma_z)/2\rangle_t=
(e^{\omega\beta}+1)^{-1}\in [0,1]$ depends on temperature, where $s=1/2$ corresponds to $T=1/\beta=\infty$ (we set the Boltzmann and Planck constant equal to one). The definition of $s$ stems from laser physics where inversion occurs corresponding to ``negative temperatures". Many authors use $\bar{n}=1/(e^{\omega\beta}-1)$ instead of $s$ and $\bar{n}+1$ instead of $(1\!-\!s)$. Then, no negative temperatures are possible. The noise channel is derived assuming certain approximations, e.g. the Markov approximation. Note, however, that this is not an essential assumption as we will demonstrate later in an example (see Figure~\ref{spingas} and related text).

An important special case of~(\ref{localnoise}), obtained by setting $B=0 $ and $C=2\gamma$, is the dephasing channel
\begin{equation}\label{dephasingchannel}
{\cal L}_{\rm{deph}}\rho=\gamma\sum_{i=1}^N\left[\sigma_z^{(i)}\rho\sigma_z^{(i)}-\rho\right],
\end{equation}
well-known especially in its integrated form as a completely positive map ${\cal E}^(i)_{\rm{deph}}(\rho)=p\rho+(1-p)/2\left(\sigma_z^{(i)}\rho\sigma_z^{(i)}+\rho\right)$ with $p=\exp\{-2\gamma t\}$. As with the Ising Hamiltonian~(\ref{Isinghamiltonian}), we will often use the dephasing channel for analytic discussions because of its simplicity.

We are interested in the steady state of this master equation for $N=2$ qubits and the question whether there is entanglement in this state. At this point we simply state the following results since we will later solve a more general master equation that contains equation~(\ref{gasME}) as a special case.

As an easy example we start with the Ising interaction Hamiltonian~(\ref{Isinghamiltonian}). The steady state will be the tensor product of the thermal states of each free Hamiltonian $H_{\rm{free}}^{(i)}=\omega/2\,\sigma_z^{(i)}$ since this state commutes with $H_{\rm{Ising}}$ and since dephasing noise does not change the diagonal elements of the density matrix. In conclusion, we have the unentangled steady state \[\rho_{\rm{steady}}=\mbox{diag}(s^2,s(1-s),s(1-s),(1-s)^2)\mbox{.}\]

When can we hope to find an entangled steady state? As we know the radiative decoherence processes drive the system into the thermal states of the free Hamiltonian. If the interaction Hamiltonian can entangle these states we may find an entangled steady state at least for low temperatures. As an example, consider an interaction Hamiltonian $H=g \sigma_x^{(1)}\sigma_x^{(2)}$ while the other terms stay the same as in the example above. We set $C=1/2\,B$ (no non-dissipative processes). The steady-state density matrix is then

\[\fl
\left(
\begin{array}{llll}
 \frac{g^2+s^2 \left(B^2+4 \omega ^2\right)}{B^2+4 \left(g^2+\omega ^2\right)} & 0 & 0 & \frac{g (2 s-1) (i B+2 \omega )}{B^2+4 \left(g^2+\omega ^2\right)} \\
 0 & \frac{g^2-(s-1) s \left(B^2+4 \omega ^2\right)}{B^2+4 \left(g^2+\omega ^2\right)} & 0 & 0 \\
 0 & 0 & \frac{g^2-(s-1) s \left(B^2+4 \omega ^2\right)}{B^2+4 \left(g^2+\omega ^2\right)} & 0 \\
 \frac{g (2 s-1) (2 \omega -i B)}{B^2+4 \left(g^2+\omega ^2\right)} & 0 & 0 & \frac{g^2+B^2 (s-1)^2+4 (s-1)^2 \omega ^2}{B^2+4 \left(g^2+\omega ^2\right)}
\end{array}
\right)
\]

This density matrix can be entangled. We measure the entanglement between two sets of qubits by the negativity~\cite{eisert_phd,vidal_negativity}, which is given with respect to a bipartition $A$-$\bar{A}$ as ${\cal N}_{A}=(||\rho^{T_{A}}||_1-1)/2$, where $T_A$ means the partial transpose with respect to $A$.
For two qubits, we omit the label $A$ since there is only one bipartition, and the negativity can assume values between zero (separable state) and $1/2$ (maximally entangled state). The reason why we choose the negativity as a measure throughout the paper is that we will use a generalization thereof in the multipartite case where the generalization of other entanglement measures might be hard to compute.

\begin{figure}[ht]
\includegraphics[width=0.7\textwidth]{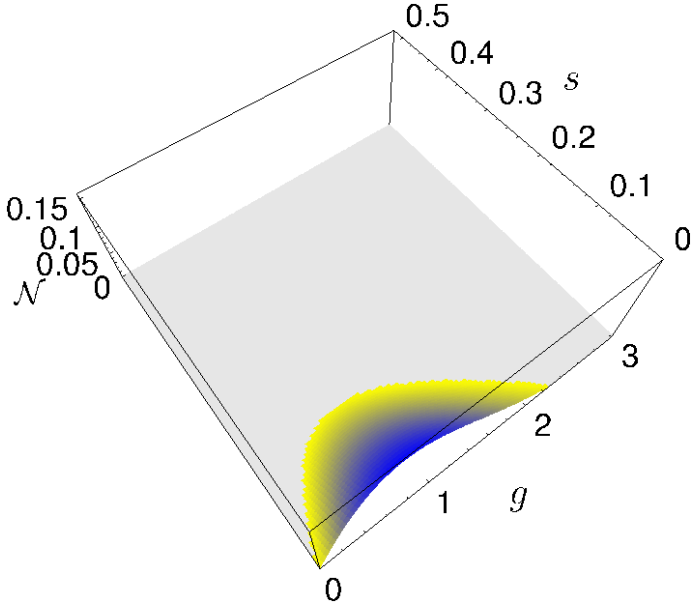}
\caption[]{\label{sxsx} Entanglement as measured by the negativity in the steady state of equation~(\ref{gasME}) with interaction Hamiltonian $H=g \sigma_x^{(1)}\sigma_x^{(2)}$ and free Hamiltonian $H_{free}=\omega/2\sum_{i=1}^N\sigma_z^{(i)}$. The decoherence processes are given by equation~(\ref{localnoise}). We choose the decay rate of inversion, $B$, as inverse unit timescale and the other parameters as $C=B/2$ (no dephasing noise) and $\omega=B$. The parameter $g$ on one of the axes is also measured in units of $B$, whereas $s$ is dimensionless.}
\end{figure}

The negativity of the state above is
\begin{eqnarray}
\fl{\cal N}=\mbox{Max}\Big\{0,\Big[\left(B^2+4 \left(g^2+\omega ^2\right)\right) \left((s-1) s \left(B^2+4 \omega ^2\right)-g^2\right)\nonumber\\
\fl\quad\quad+g |1-2 s| \left(B^2+4 \omega ^2\right)^{1/2}\left(B^2+4 \left(g^2+\omega ^2\right)\right)\Big]\left(B^2+4 \left(g^2+\omega ^2\right)\right)^{-2}\Big\},
\end{eqnarray}
which can be larger than zero but will always vanish for high temperatures of the bath, $s\to 1/2$ (see Figure~\ref{sxsx}).
We see that the steady state of the master equation~(\ref{gasME}) can be entangled for specially chosen interaction Hamiltonians, but only below a certain temperature threshold. Furthermore, if the free, local Hamiltonian is too strong, i.e., $\omega$ dominates by far all other parameters, there is also no entanglement. This statement applies to other models involving the Hamiltonian~(\ref{atomicHamiltonian}) as well, and, accordingly, $\omega$ should have the same order of magnitude as the other parameters.

\subsection{Example for a gas-type system: the spin gas}\label{prototype_spingas}

Spin gases~\cite{calsamiglia_spin_gases, spin_gases} are an example for such gas-type systems. A spin gas is a system of quantum spins with stochastic, time-dependent interactions.  A physical model of a spin gas is a system of $N$ classically moving particles with additional, internal spin degrees of freedom. Upon collision, these quantum degrees of freedom interact according to some specified Hamiltonian. In~\cite{calsamiglia_spin_gases, spin_gases} the interaction Hamiltonians were chosen locally unitarily equivalent to the Ising interaction leading to a description in terms of weighted graph states. Hence, in such spin gases, classical kinematics drives the evolution of the quantum state, and also the decoherence of arbitrary probe systems put into the gas and subjected to interactions with it. In general, multiple non-consecutive collisions of particles are possible. The spin gas remembers its whole interaction history, and it provides a microscopic model with {\em non-Markovian decoherence}.

Assume that we have two selected gas particles (e.g. another species) that we consider as the system, while the other gas particles act as the environment that induces decoherence when interacting with the system particles. The rare interactions between the two system particles happen only during the short times when they collide. In the longer times in between they are not coupled and subject to local decoherence processes, i.e., interactions with the environment gas particles. The induced decoherence processes are equivalent to dephasing channels (corresponding to $B=0, C=2\gamma$ in equation~(\ref{localnoise})).
In such a situation, any entanglement between the two qubits that may either have been present initially or have built up on a short time scale will eventually be destroyed by the interactions with the other gas particles~\cite{hartmann_reset}. \\

\noindent{\em The spin gas with reset mechanism}\\

\begin{figure}[ht]
\includegraphics[width=0.7\textwidth]{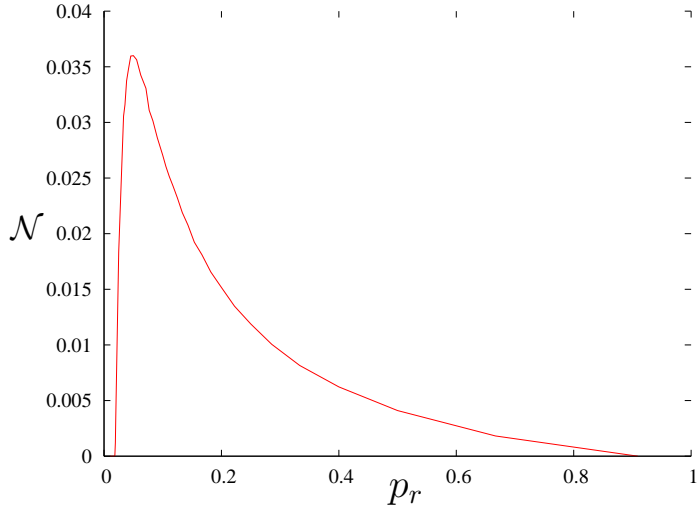}
\caption[]{\label{spingas} Steady-state entanglement between two selected qubits as measured by the negativity in a spin lattice gas ($8\times8$ lattice, $2+18$ qubits). The probability that a particle is exchanged for a fresh one in one time step of the simulation is plotted on the horizontal axes. Hence, the value $1$ corresponds to an infinite exchange rate. The special qubits interact $1000$ times stronger with each other than with the $18$ qubits that form the environment, i.e., as physical particles, they are e.g. of a different kind than the environment particles. The density matrices from which the negativity is derived were averaged over $10000$ simulation runs. Other parameters: initial distance between special qubits: $1$, interaction phase picked up during a collision between them: $\psi=0.1$; interaction phase for interactions with environment spins: $\phi=0.0001$; for details see~\cite{calsamiglia_spin_gases,spin_gases}.}
\end{figure}

For the moment, we stick to the toy-model with only to selected gas particles.
Imagine now that the two particles can, at a certain rate, leave the box in which the gas is contained and are instantly replaced by fresh qubits that are in a standard mixed state with sufficiently low entropy. Instead of a replacement of system particles one can equivalently picture a measurement of the particle and a subsequent preparation in this standard state. Note that the last step need not be an active procedure but can, e.g., result from a spontaneous decay to this state. We call both procedures a reset mechanism. Certainly, by a reset, we did not introduce entanglement into the system always consisting of two qubits. On the contrary, any entanglement that might have been present between the particle that has left the box and the one that is still inside leads to a description of the latter by a more mixed density matrix (closer to the identity). But the advantage is that we have lowered the local entropy of the system since the new particle has no correlations with the environment. This new particle can then become entangled with the other one on short time scales. We said above that in a spin gas with zero rate of qubit exchange the steady state will not be entangled. For infinite exchange rate the system would always be in a pure (or a standard separable) state and there is also no entanglement. If, however, the rate at which the qubits leave the spin gas is in a certain intermediate parameter regime one can hope that there is entanglement in the system on average. Here, averaging means taking the mean density matrix of many simulation runs. Later, the solutions to master equations are assumed to resemble the evolution of such a mean density matrix and, for explanations of certain (entanglement) features, this picture will sometimes be invoked. Note that there are also other ansatzes. In~\cite{briegel_coarse_grain} the solution of the master equation represents a smoothed version of a single simulation run, where smoothing is achieved by a time-integration kernel. The solution of the master equation does then not follow the rapid changes of the single density matrix, but sees only the slower changes resolved by a so-called coarse-grained timescale, which is related to the support of the integration kernel.
Figure~\ref{spingas} indeed shows entanglement in steady states in a simulation of a spin lattice gas with an Ising-type interaction~\cite{calsamiglia_spin_gases,spin_gases}.

The above scenario with two qubits might seem a little artificial. However, if we extend the setup to systems with more qubits and allow fluctuating particle numbers we can drop the requirement that selected particles must be instantaneously replaced. We can regard the spin-gas particles that do not belong to the system as a ``hot bath" introducing decoherence and destroying entanglement in the system, and the source of qubits in standard states as a ``cold bath" that can counteract the effect of the hot bath and preserve entanglement in the steady state. The analogy with a cold bath has some limitations as we will point out in section~\ref{stronglycoupledmodel}. We will deal with multipartite qubit systems in section~\ref{multipartite}. For the moment, we will stick to the two qubit system, for which we can find a master equation that reflects the properties of the ``cold bath" and that we can solve analytically in certain cases. Observe that the master equation will again incorporate the Markov-assumption, whereas the spin gases are non-Markovian systems~\cite{calsamiglia_spin_gases, spin_gases} and also partly have non-local decoherence processes. The essential features on the other hand will be qualitatively the same in some parameter regimes of the spin gas, where these effects play a minor role.

In the next subsection, our goal is to transfer the idea of a reset mechanism from the specific example of a spin gas to a description in form of a master equation, suitable for any spin system.

\subsection{The master equation with reset mechanism}\label{MEwithreset}

Compared to~(\ref{gasME}), the master equation that models a gas-type system with reset mechanism has an additional term ${\cal L}_{reset}$, which we describe as follows. With some probability $r\delta t$ particle $i$, $i=1...N$, is reset during the time interval $\delta t$ to some specific state $|\chi_i\rangle$. The other qubits are left in the state $\mbox{tr}_i\rho$. The change in the density matrix during the time $\delta t$ due to ${\cal L}_{reset}$ is $(\delta\rho)_{reset}=r\delta t\sum_{i=1,2}(|\chi_i\rangle\langle\chi_i|\mbox{tr}_i\rho-\rho)\equiv \delta t{\cal L}_{reset}\rho$. Observe that the time interval must be longer than the timescale of any of the involved processes but short enough so that we can replace it by the time differential to obtain, for the rate of change $\dot{\rho}=\partial \rho/\partial t$, the following master equation:
\begin{equation}\label{GeneralME}
\dot{\rho}=-i[H_{total},\rho]+{\cal L}_{noise}\rho+\sum_{i=1}^N r(|\chi_i\rangle\langle\chi_i|\mbox{tr}_i\rho-\rho)
\end{equation}

Before we proceed to discuss the solution of (\ref{GeneralME}) let us establish that the problem is well-defined, i.e., that the master equation leads to a completely positive map, which is true when the master equation is of Lindblad form. For the noise part this is known, so we have to bother only about the reset part.
Since ${\cal L}_{reset}=\sum_{j=1}^Nr(|\chi_j\rangle\langle\chi_j|\mbox{tr}_j\rho-\rho)$  is local we have to show that each summand is of the form $\sum_{m,n=1}^3 L_{mn}^j([\sigma_m^{(j)}\rho,\sigma_n^{(j)}]+[\sigma_m^{(j)},\rho\sigma_n^{(j)}])$ where the $\sigma^{(j)}$s are Pauli operators and $L^j$ must be positive (semidefinite) matrices. We expand $\rho$ and $|\chi_j\rangle\langle\chi_j|$ in the $\sigma$-basis as $\rho=\sum_{k_1,\dots,k_N=0}^3a_{k_1,\dots,k_N}\sigma_{k_1}^{(1)}\dots\sigma_{k_N}^{(N)}$ and $|\chi_j\rangle\langle\chi_j|=\sum_{q=0}^3 b_q^j\sigma_q^{(j)}$. We insert these expressions into ${\cal L}_{reset}$ and also into the Lindblad-expression, collect the coefficients that belong to each $\sigma$-matrix of $\rho$ using the scalar product, and compare the coefficients $a_{kl}$ in each expression, which leads us to a simple linear system of equations for the $L_{mn}^j$. Solving this system of linear equations we obtain
\begin{equation}\label{LindbladL}
L^j=r\left(\begin{array}{lll}\frac{1}{8}&-\frac{1}{4} i b^j_3&\frac{1}{4} i b_2^j\\ \frac{1}{4} i b_3^j&\frac{1}{8}&-\frac{1}{4} ib^j_1\\-\frac{1}{4} ib^j_2&\frac{1}{4} ib^j_1&\frac{1}{8} \end{array}\right)
\end{equation}
The eigenvalues of $L^j$ are $r/8, r/8\left(1\pm2\big((b_1^j)^2+(b_2^j)^2+(b_3^j)^2\big)^{1/2}\right)$. Since we assumed $|\chi_j\rangle\langle\chi_j|$ to be pure, we have $(b_1^j)^2+(b_2^j)^2+(b_3^j)^2=1/4$ and eigenvalues $r/8,0,r/4$. We note that also a mixed reset state would be fine to ensure that the $L^j$ are positive semidefinite. Because the sum of positive matrices is a positive matrix, and because we know that the noise terms also have positive $L$-matrices, we have shown that the master equation is of Lindblad form and preserves the positivity of the density matrix.

Up to this point, we have modeled interacting, gas-type systems coupled to a noisy environment. We have described a toy model consisting of only two particles by a master equation and compared predictions about the entanglement properties of steady states to simulations with a spin gas as an example for such gas-type systems. We have seen that in general there will be no entanglement in the steady state. We have extended the example of the spin gas by allowing particle exchange with a ``cold bath" of particles in standard states (or an equivalent reset mechanism), and we have found that steady states of such systems can be entangled. We have derived a master equation that models systems with reset mechanism and have proved that the master equation is of Lindblad form. In the following subsection we study the solutions of~(\ref{GeneralME}).

\subsection{Solution of the master equation for the gas-type model}\label{SolutionMEgastype}

In principle, the solution to the master equation~(\ref{GeneralME}) with noise channels as in equation~(\ref{localnoise}) is simple. The equation is of the form
\[
\dot{\rho}={\cal L}\rho
\]
with solution
\[
\rho(t)=e^{{\cal L}t}\rho(0)\mbox{.}
\]
Mapping $\rho$ to a column vector $C$ containing the $16$ coefficients $C_{0000},C_{0001},\dots,C_{1111}$ of the density matrix and accordingly mapping the Liouville operator ${\cal L}$ to a $16\times16$-matrix $\Lambda$ we get the equivalent $16$ coupled linear differential equations $\dot{C}=\Lambda C$ with solution $C(t)=e^{\Lambda t}C(0)$. To compute the matrix exponential we need the spectral representation of $\Lambda$, i.e., we must solve the eigenproblem $\Lambda C_\lambda=\lambda C_\lambda$. Observe that the steady state (if it exists) is given by the eigenvector $C_0$ corresponding to the eigenvalue $\lambda=0$.

In the following we will first analyze the solutions for an Ising interaction Hamiltonian and later generalize to generic cases.

\subsubsection{Ising Hamiltonian.}\label{gastype_solutionIsing}

We specialize to the Ising Hamiltonian~(\ref{Isinghamiltonian}) as (effective) interaction Hamiltonian, $H=H_{\rm Ising}$, and to a specific reset state, namely $|+\rangle\langle +|$ for both qubits. The free Hamiltonian is $H_{\rm free}=\omega/2\sum_{i=1}^N\sigma_z^{(i)}$ as before.
We can solve the problem through spectral decomposition of the Liouville operator ${\cal L}(\rho)=-i[H_{\rm total},\rho]+{\cal L}_{\rm noise}\rho+{\cal L}_{\rm reset}\rho$, but the expression for the corresponding matrix $e^{\Lambda t}$ is very lengthy.

One obtains shorter expressions if one does not solve all $16$ differential equations at once through the matrix exponential, but step by step, since not all differential equations are coupled. Still, we have chosen to move the solution derived in this way to~\ref{appendix} not to overburden the text with technical details.

For illustration, we will restrict the noise to the special case of a dephasing channel~(\ref{dephasingchannel}) in the following.\\

\noindent{\em Dephasing channel}\\

As pointed out above, the solution is given by the spectrum of the total Liouville super-operator defined by $\dot{\rho}={\cal L}\rho$ and its corresponding eigenvectors. In the case of a dephasing channel one obtains the eigenvalues
\begin{eqnarray*}
\fl\{0,-r, -2r, -2(r+2\gamma), -2(r+2\gamma\pm i\omega),\\ \fl-1/2(3r+4\gamma+\sqrt{-16g^2+r^2}\pm2i\omega,-1/2(3r+4\gamma-\sqrt{-16g^2+r^2}\pm2i\omega\}
\end{eqnarray*}
with  multiplicities $\{1,2,1,2,1+1,2+2,2+2\}$, respectively. The eigenvector belonging to the eigenvalue $0$ represents the density matrix in the steady state, and we will come back to this matrix in the next subsection. The full solution, derived by solving the differential equations in a step-by-step manner as explained above, is given in~\ref{appendix}.

To demonstrate the time-evolution of an initial density matrix governed by the master equation we plot the entanglement between the two qubits as a function of time. Note that Figure~\ref{TimeDependence} is based on the analytic solution given in~\ref{appendix}, and the plot shows how the negativities of different initial states approach the final negativity of the steady state. We choose $\gamma^{-1}$ as unit timescale (setting $\gamma=1$). The parameters $r$, $g$, $\omega$ here have the fixed values $10\gamma$, $5\gamma$, $5\gamma$, respectively. The initial states are weighted graph states~\cite{calsamiglia_spin_gases,spin_chain_long_range} with density matrix $U(\varphi)|+\rangle\langle +|U^\dagger(\varphi)$ where $U(\varphi)=\mbox{diag}(1,1,1,e^{i\varphi})$. Through the parameter $\varphi$ we can continuously tune the entanglement in the initial state from the product state $|+\rangle\langle +|$ for $\varphi=0$ to the maximally entangled, Bell-equivalent state for $\varphi=\pi$. States that are initially highly entangled are first driven into separable states before the steady-state entanglement value is approached from below. Vice versa, an initial product state gets highly entangled first, before the steady-state value for the entanglement is reached from above.
\begin{figure}[ht]
\begin{picture}(430,200)
\put(0,0){\includegraphics[width=0.6\textwidth,clip]{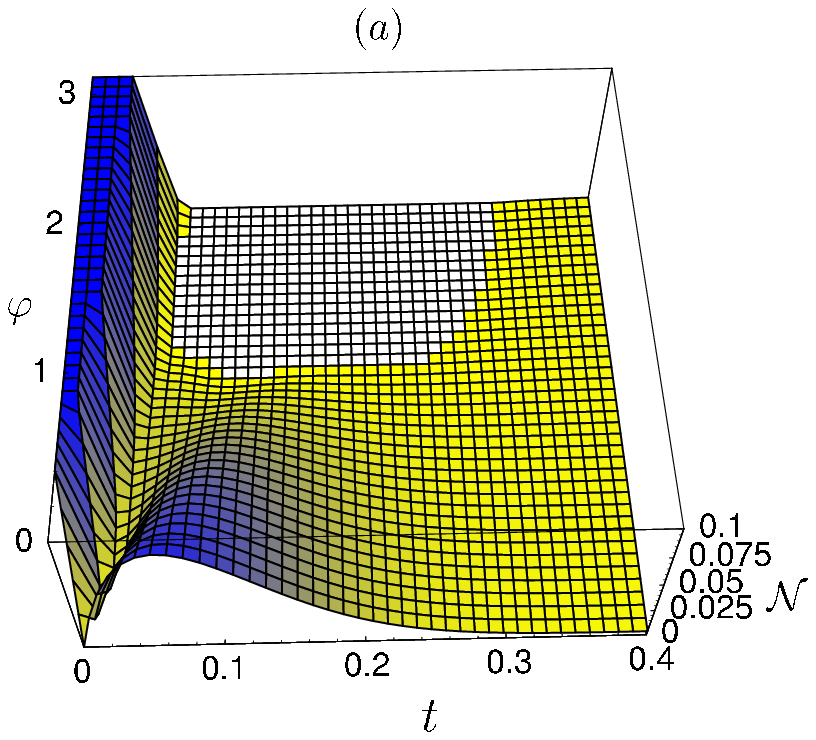}}
\put(182,65){\includegraphics[width=0.5\textwidth,clip]{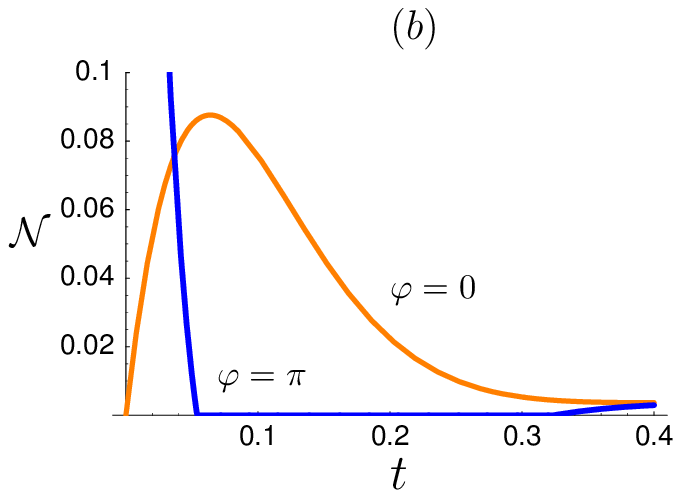}}
\end{picture}
\caption[]{\label{TimeDependence} (a) Time development of entanglement for different initial states. The $z$-axis displays the entanglement as measured by the negativity. The initial states are weighted graph states, characterized by a parameter $\varphi$ (see text). The unit timescale is $\gamma^{-1}$, $t$ is measured on this timescale, and the parameters of the master equation are $r=10\gamma$, $g=5\gamma$, $\omega=5\gamma$. (b) Cut through the $3$D plot at $\varphi=0$ (orange curve) and $\varphi=\pi$ (blue curve). }
\end{figure}
As we said earlier, to display the full analytic solution of~(\ref{GeneralME}) for more general noise channels would be quite space-consuming. We will not present it since we are primarily interested in the entanglement properties of the steady state. In the following we will discuss these properties for the master equation with general local noise channels.

\subsubsection{General steady states with Ising Hamiltonian.}

As we have seen, any initial state of the density matrix evolves exponentially fast into a steady state on a characteristic timescale given by the largest non-zero characteristic exponent (or the smallest in absolute values since they are negative). The characteristic timescale thus depends on the parameters of the master equation, too. The steady state is also a function of these parameters.

To smooth the presentation, we have again transferred the steady-state solution of~(\ref{GeneralME}) with Ising Hamiltonian, local noise channels as in~(\ref{localnoise}) and reset states $|\chi_j\rangle=|+\rangle$ to~\ref{appendix2}. Here, we illustrate the solution with a plot.

\begin{figure}[ht]
\includegraphics[width=0.7\textwidth]{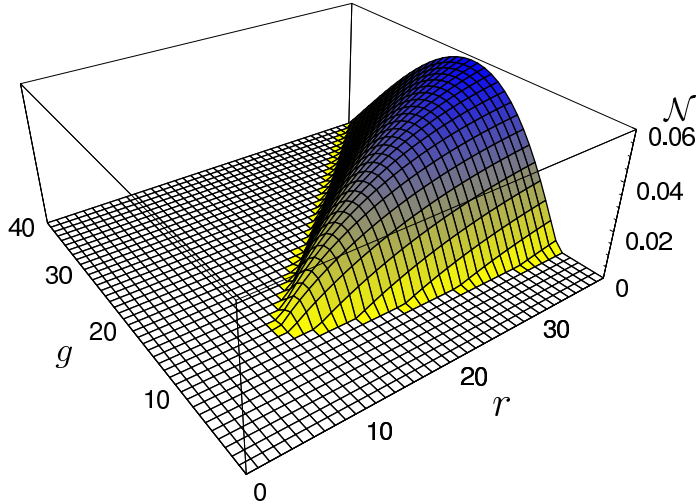}
\caption[]{\label{QOMEequState}Entanglement in the steady state of the master equation~(\ref{GeneralME}). The entanglement is measured by the negativity, the unit timescale is $B^{-1}$, the parameters $r$ and $g$ (in units of $B$) are on the axes, while the other parameters are $C=B$, $\omega=20B$, and $s=0.1$.}
\end{figure}

In Figure~\ref{QOMEequState} we choose $B^{-1}$ as unit timescale and fix the values $C=B$, $\omega=20B$ and the dimensionless parameter $s=0.1$. Then, for certain values of $r$ and $g$, measured in units of $B$, we see entanglement as measured by the negativity in the steady state.
To get entangled steady states when $g$ becomes small, we have to go to higher reset rates $r$. However, $g$ cannot be arbitrarily small. There is a weak coupling threshold below which no reset rate can ensure entanglement in the steady state. This threshold depends on the parameters of the decoherence processes, and its existence is intuitively clear. If the decoherence processes simply dominate the entangling processes, then no entanglement can be created by any means. To see this better, we momentarily put $\omega=0$ for simplification.
Let us also turn our attention once more to the dephasing channel as a special case of the noise terms of the quantum-optical master equation. Then, in the steady state, the anti-diagonal coefficients are all the same, namely $\frac{r^2(r+\gamma)}{4(r+2\gamma)(2g^2+(r+\gamma)(r+2\gamma))}$, the other off-diagonal elements are $C_{0001}=C_{0010}=C_{0111}^*=C_{1011}^*=\frac{r(-ig+r+\gamma)}{4(2g^2+(r+\gamma)(r+2\gamma))}$, and the diagonal elements all have the values $\frac{1}{4}$. All other matrix elements are given by the Hermiticity of the density matrix.
We compute from the above expressions for the density matrix the following analytic expression for the negativity in terms of the parameters $g$ (Hamiltonian interaction), $\gamma$ (strength of the dephasing channel), and $r$ (reset rate):
\begin{equation}\label{negativity}
{\cal N}
=\mbox{max}\{0,-\frac{2\gamma(r+\gamma)^2+g^2(r+2\gamma)-r(r+2\gamma)g}{2(r+2\gamma)[2g^2+(r+\gamma)(r+2\gamma)]}\}
\end{equation}
Equation~(\ref{negativity}) contains the full information about the entanglement properties of the two qubits. Note that ${\cal N}={\cal N}(\tilde g,\tilde r)$ depends, in fact, only on two parameters, $\tilde g=g/\gamma$ and $\tilde r=r/\gamma$.

\begin{figure}[ht]
\begin{picture}(230,160)
\put(0,0){\includegraphics[width=0.6\textwidth]{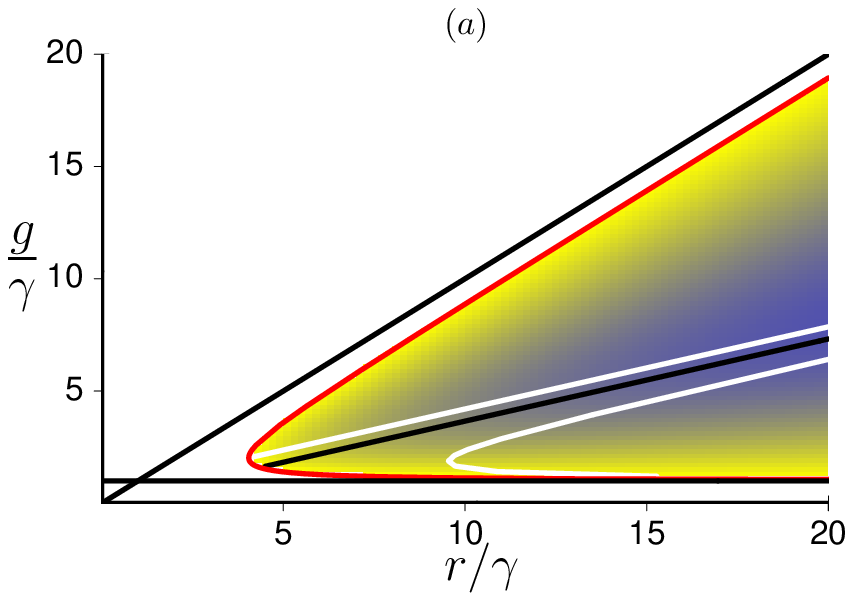}}
\put(220,47){\includegraphics[width=0.4\textwidth]{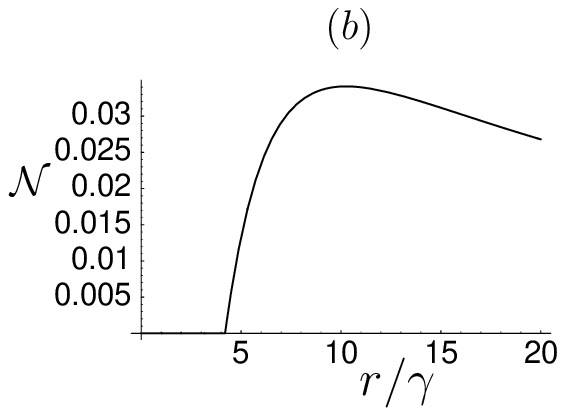}}
\end{picture}
\caption[]{\label{ToyModel} (a) Separable states (white area) and entangled states (colored area) in the $r$-$g$-plane, where $r$ is the rate of the reset process, and $g$ is the coupling strength in the Ising Hamiltonian, and we use $\gamma^{-1}=1$ as unit timescale. The color encodes the amount of entanglement measured by the negativity: the darker the area, the more entanglement is present. For a discussion of the other lines, please see the text. (b) The second plot shows a cut for constant $g=5\gamma$.}
\end{figure}

In Fig.~\ref{ToyModel}(a) we see a plot of the negativity function {\cal N}.
The key feature is the color-coded region in the $r$-$g$-plane with steady-state entanglement, where a darker color indicates higher entanglement. The entangled region is bounded by the red line given by one of the roots of the non-trivial part of equation~(\ref{negativity}). Outside of this region, the state is separable (white area). The entangled region approaches asymptotically the straights $g=2\gamma$ and $g=r$ plotted black in Fig.~\ref{ToyModel}(a). The asymptotic line $g=2\gamma$ is independent of $r$ and simply tells us that, in the weak coupling regime, decoherence/noise will always triumph over the Hamiltonian part that tries to create entanglement as pointed out before. That is, as a necessary condition, we need to be above this threshold to observe entanglement. Three lines are marked in the colored, entangled region:
\begin{enumerate}
\item The upper, white line is the maximum in $g$-direction (at constant $r$). \\
\item The lower, white line is the maximum in $r$-direction (at constant $g$).\\
\item The middle line in black is the straight $g=r/(1+\sqrt{3})$. To this middle line the upper and lower white curves go asymptotically for large $g$, $r$.
\end{enumerate}
The global maximum of the negativity is on this middle line at infinity with a value of approximately $0.0915$, about 20\% of the maximally possible value. The darkest, most entangled area in our plot has negativity approximately $0.068$.  Fig.~\ref{ToyModel}(b) shows a cut at $g=5\gamma$ through the color-plot. Most notable is the existence of a threshold value for $r/\gamma$ above which entanglement is present in the steady state.

\subsubsection{Steady state entanglement with and without reset.}

In this part we want to compare steady-state entanglement that is due to the reset to entanglement that is due to a special combination of interaction Hamiltonian and decoherence process as in subsection~\ref{MEwithoutreset}. There, the interaction Hamiltonian is
\[
H=g \sigma_x^{(1)}\sigma_x^{(2)}
\]
while the free Hamiltonian are given by~(\ref{atomicHamiltonian}). The decoherence process~(\ref{localnoise}) is determined by the parameter choice $C=B/2$. In~\ref{MEwithoutreset}, it has been established that the steady state is entangled, but only in a finite temperature range.

If we add the reset mechanism to the master equation of this example we can show that the steady state is entangled for arbitrary temperatures. As reset states we choose the eigenstate $|1\rangle=-\sigma_z|1\rangle$ of the Pauli operator $\sigma_z$.

The steady-state density matrix of the master equation has now the matrix elements
\begin{eqnarray}\label{ESwithandwithoutr}
\fl C_{0000}=\frac{B^2 s^2 \omega ^2+(B+2 r) \left((B+r) g^2+B^2 (B+2 r) s^2\right)}{(B+r) \left((B+r) \omega ^2+(B+2 r) \left(4 g^2+(B+r) (B+2 r)\right)\right)}\nonumber\\
\fl C_{0101}=C_{1010}=\left\{(B+r) \left((B+r) \omega ^2+(B+2 r) \left(4 g^2+(B+r) (B+2r)\right)\right)\right\}^{-1}\nonumber\\
\fl\times\left\{(B+2 r) \left((B+r) g^2-B^2 (B+2 r) s^2+B (B+r) (B+2 r) s\right)-B s (s B-B-r) \omega ^2\right\}\nonumber\\
\fl C_{1111}=\left\{(B+r) \left((B+r) \omega ^2+(B+2 r)\left(4 g^2+(B+r) (B+2 r)\right)\right)\right\}^{-1}\nonumber\\
\fl\left\{(-s B+B+r)^2\omega^2\right.\nonumber\\
\fl\left.+(B+2 r)\left[B^2 (B+2 r) s^2-2 B (B+r) (B+2 r) s+(B+r) \left(g^2+(B+r) (B+2 r)\right)\right]\right\}\nonumber\\
\fl C_{0011}=\frac{g (2 s B-B-r) (i (B+2 r)+\omega )}{(B+r) \omega ^2+(B+2 r) \left(4 g^2+(B+r) (B+2 r)\right)}
\end{eqnarray}
while all other coefficients are zero or are given by Hermiticity.
The negativity of this density matrix is
\begin{eqnarray}\label{NEGwithandwithoutr}
\fl{\cal N}
=\mbox{max}\left\{0,-\frac{1}{4} \left(-\frac{\left((B+2 r)^2+\omega ^2\right) (-2 s B+B+r)^2}{(B+r) \left((B+r) \omega ^2+(B+2 r) \left(4 g^2+(B+r) (B+2 r)\right)\right)}\right.\right.\nonumber\\
\left.\left.-4\sqrt{\frac{g^2 (-2 s B+B+r)^2 \left((B+2 r)^2+\omega ^2\right)}{\left((B+r) \omega ^2+(B+2 r) \left(4 g^2+(B+r) (B+2r)\right)\right)^2}}+1\right)\right\}
\end{eqnarray}

\begin{figure}[ht]
\begin{picture}(430,200)
\put(0,0){\includegraphics[width=0.55\textwidth]{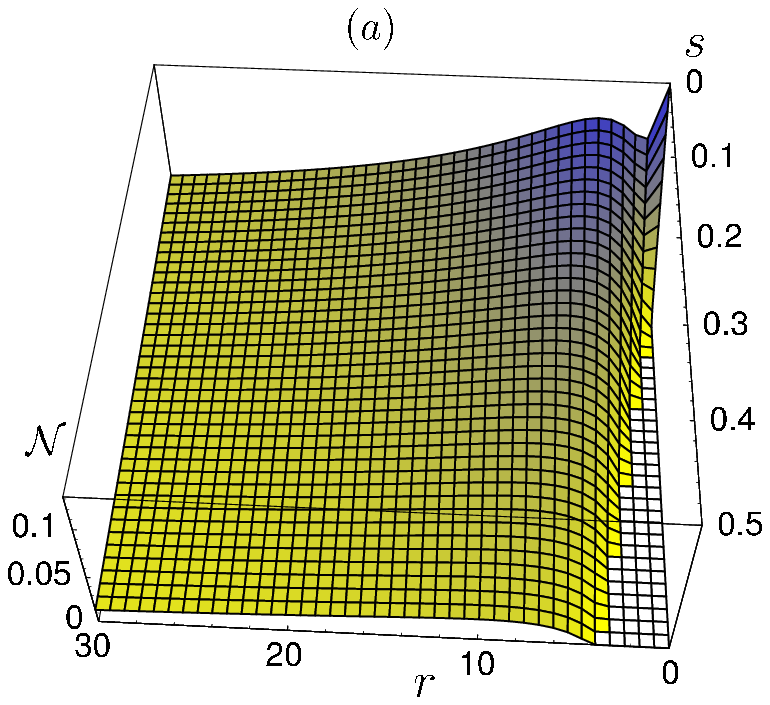}}
\put(204,38){\includegraphics[width=0.45\textwidth]{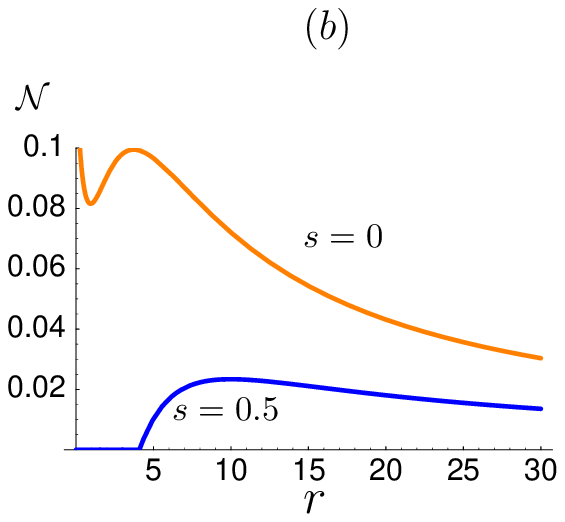}}
\end{picture}
\caption[]{\label{variationOfsAndr}(a) Negativity for $2$-qubit system with $H_{total}=g\sigma_x^{(1)}\sigma_x^{(2)}+\omega/2(\sigma_z^{(1)}+\sigma_z^{(2)})$ as a function of the reset rate $r$ and temperature-dependent parameter $s$. The noise is described by the quantum-optical master equation~(\ref{localnoise}) with $B=2C=1$. With $B^{-1}$ as unit timescale the other parameters are given by $g=2B$ and $\omega=2B$. The entanglement due to the reset mechanism exists for all temperatures, while the entanglement without reset mechanism ($r=0$) vanishes above a certain temperature threshold. (b) Cut at constant $s=0$ (orange curve, corresponding to zero temperature) and $s=0.5$ (blue curve, corresponding to infinite temperature).}
\end{figure}

We plot the function~(\ref{NEGwithandwithoutr}) in Figure~\ref{variationOfsAndr}; see figure caption for details. We observe two features:
\begin{enumerate}
\item At $r=0$ we are back to the situation of subsection~\ref{MEwithoutreset}, where entanglement vanishes above some temperature threshold (remember: $T\rightarrow\infty$ as $s\rightarrow1/2$).\\
\item There is a threshold reset rate $r$, above which an entangled steady state exists for arbitrary temperature.
\end{enumerate}
From the coefficients of the steady-state density matrix~(\ref{ESwithandwithoutr}) we also see that the entanglement created by the reset stems from a different density matrix than the entanglement present without reset. In Figure~\ref{variationOfsAndr} this is visible in the region of small $r$, where the reset tends to destroy this latter entanglement. Then, for larger $r$, the effect of the reset mechanism kicks in.

Up to now we studied rather special interaction Hamiltonians. In the following we demonstrate the genericity of entanglement that is present in a steady state due to a reset mechanism.

\subsubsection{Generic cases.}

In the previous example we have pointed out that entanglement, if present at all without reset, stems from a special combination of Hamiltonian and decoherence process. One may ask whether adding a reset mechanism with special reset states is not just as artificial as the choice of special combinations of Hamiltonians and decoherence processes. We are now going to show that one and the same reset with fixed reset states can lead to steady-state entanglement for many combinations of Hamiltonians and decoherence processes. The reasoning was the opposite without reset, where only very few combinations of Hamiltonians and decoherence processes lead to steady-state entanglement. We can also relax the condition that the reset states are pure states to a certain extent. Hence, we show that we have found generic features by generalizing the system in various directions.

\begin{figure}[ht]
\begin{picture}(430,200)
\put(0,0){\includegraphics[width=0.55\textwidth]{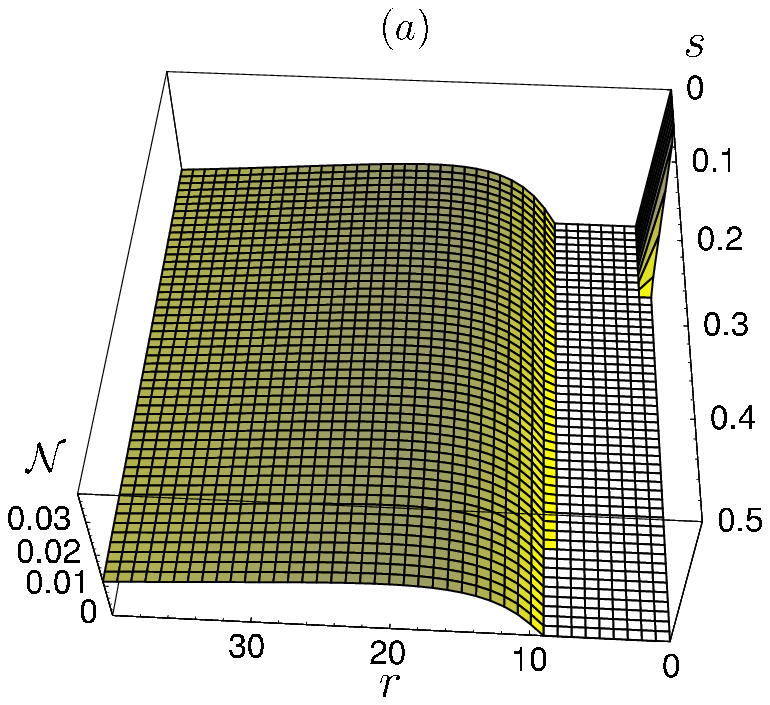}}
\put(204,38){\includegraphics[width=0.45\textwidth]{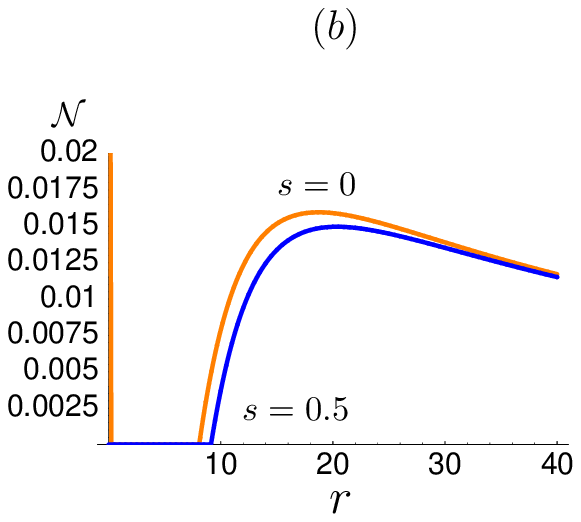}}
\end{picture}
\caption[]{\label{variationOfs}(a) Negativity for two-qubit system with $XYZ$ interaction and magnetic field, $H=g(0.7\sigma_x^{(1)}\sigma_x^{(2)}+0.3\sigma_y^{(1)}\sigma_y^{(2)}+\sigma_z^{(1)}\sigma_z^{(2)}+0.5(\sigma_x^{(1)}+\sigma_x^{(2)}))+\omega/2(\sigma_z^{(1)}+\sigma_z^{(2)}))$, as a function of the reset rate $r$ and temperature-related parameter $s$. The noise is described by the quantum-optical master equation channel~(\ref{localnoise}) with $B=2C=1$ ($B^{-1}$ as unit timescale). The Hamiltonian parameters are $g=2.5B$, $\omega=4B$. (b) Cut through the plot. The upper curve corresponds to zero temperature $(s=0)$, the lower one to infinite temperature $(s=1/2)$ of the bath. Curves for any finite temperature lie in between.}
\end{figure}

(i) The qualitative behavior of the two-qubit model does not depend on the particular choice of the interaction Hamiltonian or details of the decoherence model other than its local action on individual qubits.
Figure~\ref{variationOfs} shows e.g. steady-state entanglement for an XYZ Hamiltonian as function of reset rate $r$, and decoherence described by the noise operator ${\cal L}_{\rm noise}$. The qualitative behavior is similar to Figure~\ref{ToyModel}(b) or Figure~\ref{variationOfsAndr}, and we observe steady-state entanglement even for infinite temperature of the bath.

\begin{figure}[ht]
\includegraphics[width=0.6\textwidth]{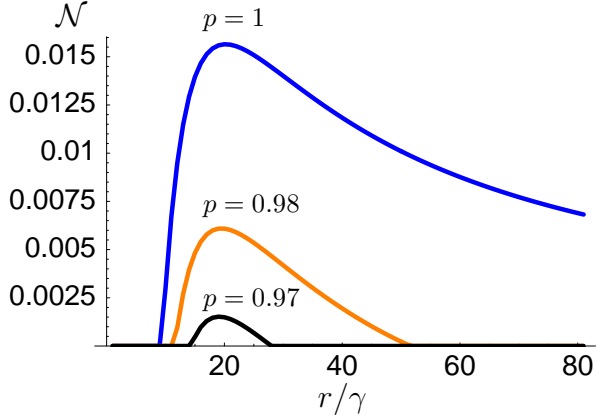}
\caption[]{\label{variationOfpImperfectprojection}Negativity for two-qubit system with $XYZ$ interaction and parameters as in Figure~(\ref{variationOfs}). The temperature related parameter has the fixed value $s=0.1$. Now, the reset state is not the pure state $\rho_{\rm reset}=|++\rangle\langle++|$ but the mixed state $\rho_{\rm reset}=p|++\rangle\langle++|+(1-p)/2 \one$. The three curves, from top to bottom, correspond to $p=1$, $p=0.98$, and $p=0.97$. With a mixed reset state, a third threshold for large but finite reset rate $r$ appears above which the state is not entangled in contrast to the case of pure reset states where the entanglement vanished only for $r\rightarrow\infty$ (topmost curve).}
\end{figure}

(ii) The idealized reset mechanism we consider can be replaced by a more realistic imperfect reset mechanism. In this case, fresh particles are in mixed states with sufficiently low entropy rather than in pure states (with entropy 0). 
Still, the steady state turns out to be entangled. When we vary $r$ there is a new, third threshold. First, for very small $r$, there can be an entangled steady state, which is not due to the reset and which is present only for a finite temperature range above zero. Second, there is one threshold value for $r$ above which the steady state is entangled due to the reset mechanism (for arbitrary temperature). Third, whereas for pure reset states the entanglement goes down to zero again only in the limit $r\rightarrow\infty$ (permanent projection to the product of the reset states), the entanglement goes down to zero for some finite $r$ if the reset states are mixed states. This behavior is easy to understand, since it is more difficult for the interaction Hamiltonian to create entangled states from mixed reset states. The higher the entropies of the reset states are, the smaller is the range of the reset rate $r$ for which there is entanglement in the steady state. This range can also become zero, so we must demand reset states of sufficiently low entropy. Figure~(\ref{variationOfpImperfectprojection}) clearly shows this new threshold appearing for large reset rates.

The picture that emerges from all these results is the following.
Entanglement can prevail in dissipative, open quantum systems that are far away from thermodynamic equilibrium. For gas-type systems treated in this section, a reset mechanism can evoke steady-state entanglement even for infinite temperature of the environment generically, i.e., independently of the specific form of the interaction Hamiltonian or decoherence channel.

In the next section we show that steady-state entanglement appears also in strongly coupled systems with an appropriate reset mechanism.

\section{Strongly coupled systems}\label{stronglycoupledmodel}
In gas-type systems, we can treat the local noise channels separately for each qubit as explained above. If these local channels correspond to a heat bath, they drive each qubit individually to the thermal state of the local, free Hamiltonian, i.e., they populate the eigenstates of the free Hamiltonian according to the Boltzmann factor. Although the effective interaction Hamiltonian in the master equations is represented as continuously acting, the physical interaction process in gas-type scenarios is viewed as a short collision event. Hence, the interaction Hamiltonian does not influence the energy spectrum considerably and does not modify this thermal state. In strongly coupled systems, on the other hand, interactions of quanta of the heat bath with the system qubits affect the system as a whole. In this sense, the decoherence process acts globally on the system, inducing transitions between joint eigenstates. In this section we will shortly discuss the master equation describing a strongly coupled spin system in contact with a thermal, photonic bath. We will see that the resulting equilibrium state, the thermal state, can be entangled below a certain temperature threshold if the ground state of the Hamiltonian is entangled. When we add a reset mechanism we find that, in contrast to the gas-type scenario, entanglement in the steady state can exist only below a certain temperature threshold. However, the novel feature is that this threshold is typically much higher than for the thermal state. Finally, we describe the influence of the master equation parameters on the respective steady state, and, with this insight, formulate a general condition under which a reset mechanism can lead to an entangled steady state. \\

Let $|a\rangle$ ($|b\rangle$) be momentary eigenstates of some non-degenerate system Hamiltonian $H(t)$ with eigenenergies $\omega_a$ ($\omega_b$)\footnote{Note that $\hbar=k_B=1$ throughout the paper.}. We define $N_{ab}:=(e^{\beta(\omega_a-\omega_b)}-1)^{-1}$ with $\beta=1/T$ being the inverse temperature. Often $N_{ab}$ is written as $\bar{n}$, and we explained the connection to the parameter $s$ in the last section. The master equation for a spin system, coupled with strength $\gamma$ to a heat bath consisting of photons, is~\cite{childs_adiabatic_qc}

\begin{eqnarray}\label{MEsolidstate}
\fl\dot{\rho}=-i[H,\rho]-\gamma\sum_{j,a,b}[N_{ba}|g_{ba}|^2|\langle a|\sigma_-^{(j)}|b\rangle|^2+(N_{ab}+1)|g_{ab}|^2|\langle b|\sigma_-^{(j)}|a\rangle|^2]\nonumber\\
\qquad\times\left\{|a\rangle\langle a|\rho+\rho|a\rangle\langle a|-2\langle a|\rho|a\rangle |b\rangle\langle b|\right\},
\end{eqnarray}
Here, $g(\omega)$ is the spectral density, for which $g_{ba}= g(\omega_b-\omega_a)$ if $\omega_b>\omega_a$ and $g_{ba}=0$ else. For small system one can justify to treat the spectral density as constant ($g=1$ if $\omega_b>\omega_a$) and merely tune the overall coupling constant $\gamma$~\cite{childs_adiabatic_qc}. Observe that we did not include non-radiative contributions as opposed to the master equation~(\ref{gasME}) with noise terms~(\ref{localnoise}).

The master equation~(\ref{MEsolidstate}) drives any initial density matrix to the thermal state of inverse temperature $\beta$. That means, the ground state and also the excited states are populated according to the canonical distribution.

We will study the master equation~(\ref{MEsolidstate}) for an Ising Hamiltonian with transverse magnetic field, briefly discuss the solution without reset mechanism (thermal state), and then turn to an analysis of the full master equation with reset mechanism. 

\subsection{Master equation without reset}

An Ising Hamiltonian with transverse magnetic field has the form
\begin{equation}\label{HIsingTransvMagnField}
H_I=g[\sigma_z^{(1)}\sigma_z^{(2)}+b(\sigma_x^{(1)}+\sigma_x^{(2)})],
\end{equation}
and the eigenvalues are $-g\sqrt{1+4b^2}$,$-g$,$g$,$g\sqrt{1+4b^2}$ with corresponding eigenvectors, expressed in the standard basis,
\begin{eqnarray}\label{eigenvectorHIsing+B}
\ket{\psi_0}=N(1,(-1+\sqrt{1+4b^2})/2b,(-1-\sqrt{1+4b^2})/2b,1),\nonumber\\
\ket{\psi_1}=1/\sqrt{2}(0,-1,1,0),\nonumber\\
\ket{\psi_2}=1/\sqrt{2}(-1,0,0,1),\nonumber\\
\ket{\psi_3}=N(1,(-1-\sqrt{1+4b^2})/2b,(-1-\sqrt{1+4b^2})/2b,1),
\end{eqnarray}
where $N=(2+1/2|(-1+\sqrt{1+4b^2})/b|^2)^{-1/2}$ provides normalization.
We exclude the case $b=0$ where the ground state would be degenerate, a case not properly described by the master equation~(\ref{MEsolidstate}). The ground state of this system is the first eigenvector in equation~(\ref{eigenvectorHIsing+B}). Since this state is entangled, so is the thermal state below a certain temperature threshold. We see this directly from the negativity of the thermal state $\rho_{\rm thermal}=\exp\{-\beta H\}/\tr{\exp\{-\beta H\}}$, which is
\begin{equation}\label{strongly_coupled_neg_thermal_state}
{\cal N}(\rho_{\rm thermal})=\mbox{Max}\left\{0,-\frac{\cosh (g \beta )-\frac{\sinh \left(\sqrt{4 b^2+1} g \beta \right)}{\sqrt{4 b^2+1}}}{2 \left(\cosh (g \beta )+\cosh \left(\sqrt{4
   b^2+1} g \beta \right)\right)}\right\}.
\end{equation}
For any fixed $g$ and $b$, the non-trivial part in this formula goes to the value $-1/4$ when $\beta\rightarrow0$, while it goes to $1/(2\sqrt{4 b^2+1})>0$ for $\beta\rightarrow\infty$. The threshold value for $\beta$, where the non-trivial part becomes exactly zero, can be easily computed numerically for any given parameters $g$ and $b$. In terms of $\beta$, the thermal mixture of the eigenstates is separable below this threshold, which, in terms of $T=\beta^{-1}$, means that the mixture is separable above that critical temperature. From~\ref{strongly_coupled_neg_thermal_state} one can see that the critical temperature grows linearly with $g$ and monotonously, but sub-linearly with $b$.

\subsection{Master equation with reset}

We keep the Ising Hamiltonian with transverse magnetic field as above, but extend the master equation~(\ref{MEsolidstate}) by the reset term
\begin{equation}\label{reset_strongly_coupled_systems}
\sum_{i=1}^N r(|\chi_i\rangle\langle\chi_i|\mbox{tr}_i\rho-\rho)
\end{equation}
with $N=2$ and reset state $|\chi_i\rangle=|+\rangle$ for both qubits. We solve the resulting master equation numerically. In Figure~\ref{timedepSolidstate} we see how the entanglement for different reset rates $r$ develops over time $t$ from the value zero in the initial product state $\rho=|++\rangle\langle++|$ to its final value in the steady state while all other parameters are kept fixed (see figure caption). We notice that small, non-zero reset rates decrease the entanglement in the steady state until it is gone, while larger rates can bring entanglement back.

\subsubsection{Influence of the parameter $r$.}

\begin{figure}[ht]
\includegraphics[width=0.95\textwidth]{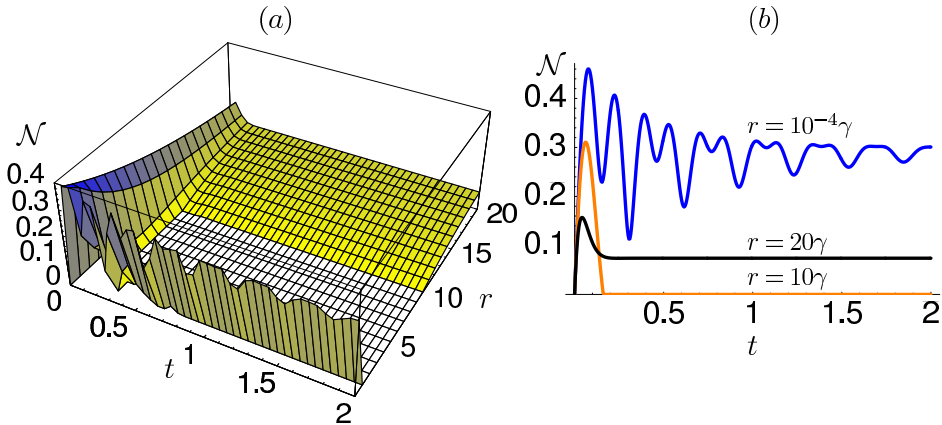}
\caption[]{\label{timedepSolidstate} (a) Solution of the master equation~(\ref{MEsolidstate}) with reset term~(\ref{reset_strongly_coupled_systems}) for a strongly coupled system. The unit timescale is $\gamma^{-1}$ and the time $t$ is plotted in these units. The temperature parameter was chosen as $\beta=1000$, and, at this low temperature, the steady state of~(\ref{MEsolidstate}) is entangled even without reset, the other parameters being $g=10\gamma$ and $b=0.1$. When increasing the parameter $r$ of the reset mechanism, the steady-state density matrix changes, as explained in the text, fist becoming separable and then entangled again. (b) Cut through the same plot for different $r$.}
\end{figure}

To explain this effect, recall that the master equation mimics the averaged density matrices that would be obtained from (infinitely many) simulation runs of the system. The reset rates of the master equation are related to probabilities that in a simulation a reset took place during a certain time interval. Although our reset processes are strictly speaking local, let us assume for the sake of argumentation that the reset happens on both qubits simultaneously, thus effectively restarting the process again from the beginning whenever a reset occurs in a simulation. For small rates $r$, i.e., for small probabilities that a reset takes place in the simulation, the system can come close to its thermal equilibrium state before it is reset. When we average the density matrices over many simulation runs, we average matrices that are mostly close to the unique thermal equilibrium state, and hence also the mixture will still retain entanglement. When the rates get larger, the density matrices over which we average become more and more diverse since they will be far from equilibrium and fluctuations occur. As a consequence the average density matrix will have no entanglement. When the reset rate is above a certain threshold, we will find entanglement in the system again (as we did in the gas-type systems) because now the density matrices over which we average become similar again. Now, they are close to the state that has unitarily evolved for a time of order $1/r$ from the initial reset state. In the limit $r\rightarrow\infty$ the state is constantly kept in the initial product state with zero entanglement. In this way we can understand how the two entangled regions arise. The first is an artifact of the entangled thermal state that is more and more destroyed by the reset mechanism. This first region could also be present in a gas-type model. The second region is the one that is really created by the reset mechanism just as in the gas-type model. We can directly see in Figure~\ref{timedepSolidstate} that entanglement in a thermal state is a truly different effect from entanglement that is created by the reset mechanism.

\subsubsection{Influence of the parameter $\gamma$.}

\begin{figure}[ht]
\includegraphics[width=0.95\textwidth]{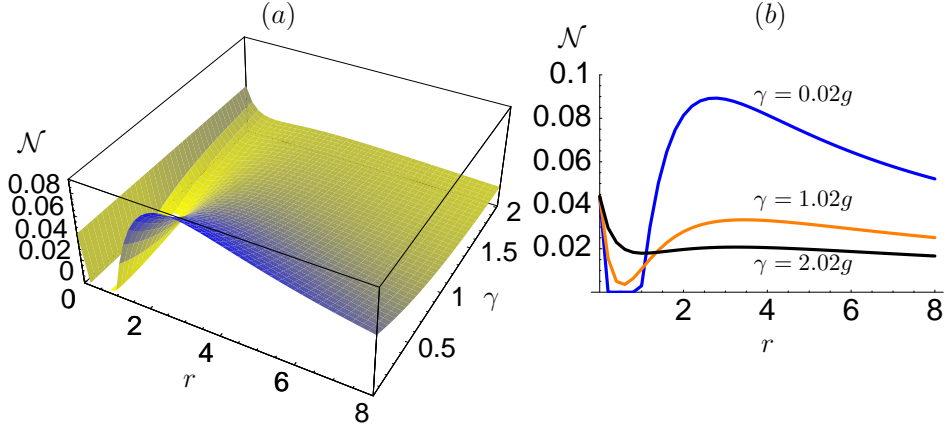}
\caption[]{\label{couplingToBath} (a) Influence of the parameter $\gamma$, which describes the strength of the coupling to the photon bath. Here, we choose $g^{-1}$ as unit time. The other parameters are $\beta=10$ and $b=0.1$. (b) Cut through the same plot for different values of $\gamma$.}
\end{figure}

Although the two effects are truly different, this does not mean that for certain parameter regimes, the two regions cannot overlap, see Figure~\ref{couplingToBath}. Imagine the coupling to the photon bath, $\gamma$, is increased. This means that the system is driven towards thermal equilibrium faster than before. Hence, following the arguments from above, the system can tolerate higher reset rates before the entanglement in the first region is destroyed. The stronger coupling to the photon bath suppresses the entanglement in the second region, and as an overall effect we see that the two regions need not be separate. Note that the entanglement in the thermal state for $r=0$ is independent of $\gamma$ because then it does not matter how fast equilibrium was approached.

\subsubsection{Influence of the parameter $b$.}

\begin{figure}[ht]
\includegraphics[width=0.98\textwidth]{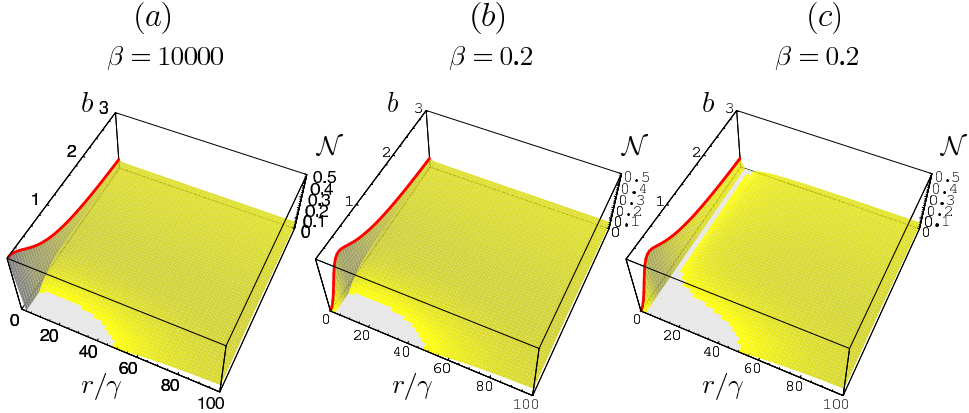}
\caption[]{\label{NvsbR} Influence of the relative magnetic field $b$. The unit timescale is given by $\gamma^{-1}$ and $g=50\gamma$. Plots (a) and (b) show the difference between (almost) zero temperature ($\beta=10000$) and some finite temperature ($\beta=0.2$). The behavior of the red curves, for $r=0$, are explained in the text. Plots (b) and (c) demonstrate the influence of different reset states in connection with the magnetic field $b$. Plot (b) has reset state $|+\rangle$, (c) has reset state $|-\rangle$.}
\end{figure}

The transverse magnetic field with relative strength $b$ splits up the energy levels of the Hamiltonian (we exclude the degenerate point $b=0$). The ground state will contain less and less entanglement as $b$ increases (and will approach the product state $|--\rangle$ for $b\rightarrow\infty$). Hence, at zero temperature, the entanglement in the equilibrium state for $r=0$ will go down for increasing $b$ (see red line in Figure~\ref{NvsbR}(a)). For a thermal state with $T\neq0$, i.e., $\beta\neq\infty$, the ground and exited states get mixed.  When $b$ is small, the splitting between ground and first exited state is small, and the mixture will be close to a separable state. For increasing $b$, the larger energy split leads to an increased population of the ground state relative to the exited states at the same temperature $1/\beta$, and the entanglement will increase. When $b$ gets even larger, the thermal state will be close to the ground state, but we know, that the ground state for large $b$ is only weakly entangled. Hence, there will be some $b$ for which the entanglement in the thermal state is maximal (see red line in Figure~\ref{NvsbR}(b), and the identical line in Figure~\ref{NvsbR}(c)). When we switch on the reset mechanism, we see that an increasing $r$ destroys the entanglement in the thermal state as before, but for increasing $b$ the regions one and two (artifact of thermal state vs. true reset state entanglement) quickly overlap. That is because the magnetic field tends to drive the state towards $|--\rangle$, whereas an increasing reset mechanism drives the state towards $|++\rangle$. If we choose the reset state as $|-\rangle$, the two effects do not compete and both drive the state towards an unentangled product state (see light grey areas in Figure~\ref{NvsbR}(c) and compare to the light grey areas in Figure~\ref{NvsbR}(b)). How fast an increasing $r$ destroys the entanglement in the thermal state is almost independent of the entanglement in the thermal state. For larger $r$ the influence of $b$ plays less and less a role and there is almost no difference between Figures~\ref{NvsbR}(b),(c) for large $r$.

\subsubsection{Influence of the parameters $g$ and $\beta$.}

\begin{figure}[ht]
\includegraphics[width=0.98\textwidth]{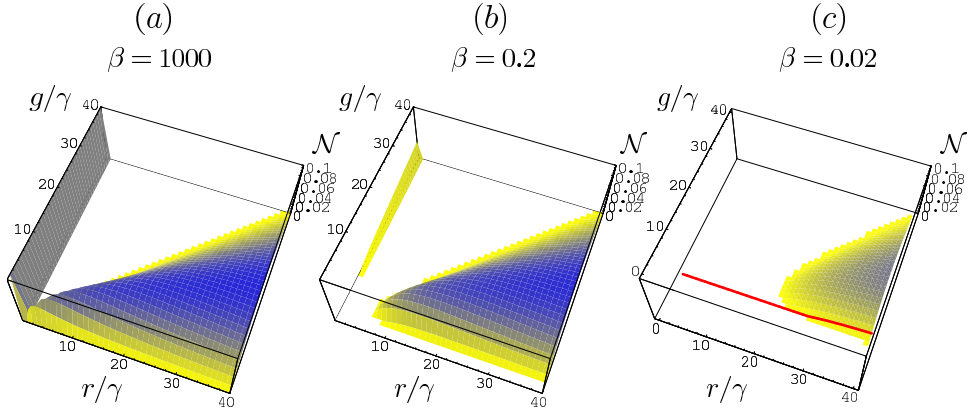}
\caption[]{\label{ESsolidstate} Negativities for increasing temperature (decreasing $\beta=1/T$). The unit timescale is given by $\gamma^{-1}$, the relative magnetic field by $b=0.1$. The parameter region in the $r-g$ plane, where steady-state entanglement occurs, becomes smaller for increasing temperatures. However, the temperatures, for which entanglement can exist with reset mechanism, are much higher than the temperatures, for which the thermal state is entangled. The red line in (c) corresponds to the red line in Figure~\ref{entCondition} (see that figure caption and the text) and is drawn for comparison only.}
\end{figure}

The coupling strength of the Hamiltonian $g$ also splits the energy levels. Hence with increasing $g$ there will be more entanglement in the thermal state at some finite temperature. Again, the speed with which an increasing reset rate destroys the entanglement in the thermal state is almost independent of $g$ (see Figure~\ref{ESsolidstate}(a)). Most interesting is the influence of the temperature $1/\beta$. Figures~\ref{ESsolidstate}(a)-(c) show plots which contain information about the equilibrium-state entanglement for different temperatures ($r=0$). We see how the thermal states get less and less entangled for increasing temperatures, so that region one vanishes quickly as expected. But, for the same temperatures, the reset rate $r$ can still produce entanglement in a steady state! On the other hand, for fixed $g$, there is some temperature threshold above which no reset rate can produce entanglement. This threshold is in contrast to the case of gas-type systems.

\subsection{General conditions for steady-state entanglement}

This discrepancy between gas-type and strongly interacting systems raises the deeper question: What are the conditions under which the reset mechanism can create entanglement in the steady state? We already see that the reset mechanism is different from a cold bath -- equivalently the replacing of system particles by fresh, standard ones -- because we would expect that some cold bath can always counteract the influence of the hot photon bath. The condition that the solution of the master equation (a completely positive (CP) map) at $r=0$ is entangling at some point in time is certainly necessary, since, as pointed out before, the reset mechanism does not introduce entanglement itself. 
The question is: Is a solution of the master equation at $r=0$ that creates entanglement on some short time scale sufficient such that the solution for some $r>0$ is a CP-map with entangled steady state? Unfortunately this is not true. The reset rate $r$ itself can influence the solution of the master equation in such a way that although the solution for $r=0$ was entangling on some short time scale, the solution for larger $r$ need not be.

\begin{figure}[ht]
\includegraphics[width=0.7\textwidth]{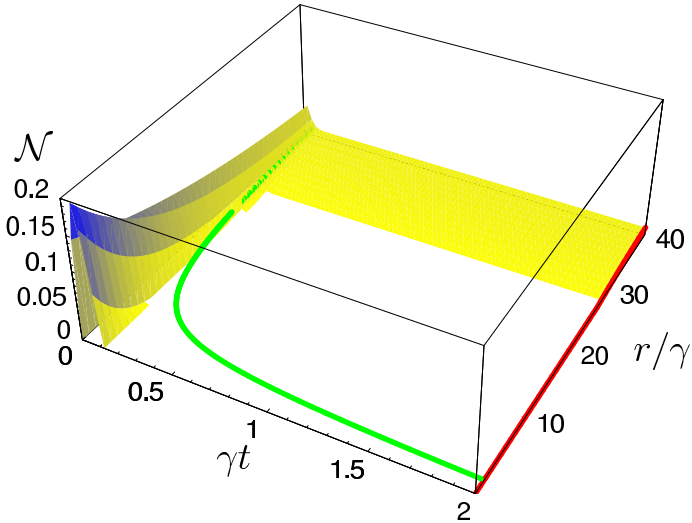}
\caption[]{\label{entCondition} Illustration of the condition under which a reset mechanism can create entanglement in the steady state. At time $\gamma t=2$ the state is already close to the true steady state, so the red line corresponds to the red line in Figure~\ref{ESsolidstate}(c) since $g=10 \gamma$ in this plot, and the other parameters are the same as in Figure~\ref{ESsolidstate}(c). The reset can create entanglement in the steady state if, for given $r$, there is entanglement in the time-evolved state at time $t\propto 1/r$ (see text for details). The green curve is the curve $t=2/r$ illustrating this result.
}
\end{figure}

The condition for steady-state entanglement is the following. If the solution of the master equation for some $r>0$ is entangling on a time scale of order $1/r$ then the steady state of this solution will also be entangled. For illustration, we think once more in terms of many hypothetical simulation runs. As stated above, the states over which we have to average will be close to the state that has unitarily evolved for a time $1/r$ from the initial reset state. When this state has a certain amount of entanglement, then so does the mixture of states close to it (see Figure~\ref{entCondition}).

The condition does not only hold for the specific Hamiltonian or the specific heat bath chosen here. It is valid for a large class of Hamiltonians (with appropriately chosen reset states) and baths.

In the next section we treat the multipartite case.

\section{Multipartite case}\label{multipartite}

For multipartite spin systems, we will show that the reset mechanism can create steady-state entanglement in a similar way. The parameter regions where this happens are comparable to the $2$-qubit case. The values of the negativity, or rather its generalization, the average negativity, stay almost constant with increasing system size. Note, however, that larger systems could have larger negativities, so if we divide the actual (constant) negativity by the maximal possible negativity, then this quantity would go down for growing system size.
We will also consider entanglement in reduced density matrices. Since in a reduction from $N$ to, say, $2$ qubits the traced out $N-2$ qubits act as an additional noise source, it is not surprising that the parameter region where we find steady-state entanglement in the reduced systems shrinks with growing $N$. If the number of particles fluctuates according to some distribution, our best description of a reduced density matrix is a mixture of reductions originating from different system sizes. It is remarkable that even in this case there is some parameter region, where steady-state entanglement is found.

In the following, we motivate and explain the entanglement measures we are using and then demonstrate the above features in both gas-type systems and strongly interacting systems.

\subsection{Gas-type systems}\label{gastypemultipartite}

It is straightforward to generalize the master equation for the gas-type system, equation~(\ref{GeneralME}) to $N>2$ qubits. The relevance of the entanglement quantities we are going to use needs to be motivated, though. We turn once more to the example of a spin gas. Imagine that the spin gas is in a box of volume $V$. This box has one semi-permeable wall through which particles can leave the box, and another through which particles from a ``cold reservoir" can enter. By ``cold reservoir" we mean that the quantum state of the particles in this reservoir is in some sufficiently pure standard state. The motional degrees of freedom of theses particles, on the other hand, are in thermal equilibrium with the outside environment just like the system particles in the box. Assume that the density of the gas in the box is $\eta$. Then there are on average $\eta V=:\lambda$ particles in the box. The distribution of the number of particles that are in the box is a Poissonian $p_\lambda(n)=e^{-\lambda}\lambda^n/n!$. When we observe the spin gas after certain time intervals, which should be long enough such that the gas always reaches its equilibrium state, we sample the distribution and get information about the density matrices with a corresponding number $n$ of qubits. The density matrices we can reconstruct after we collected a certain, sufficient amount of information is close to the steady-state density matrix of the master equation for $n$ qubits. We are interested in the entanglement properties of the gas and we will look at different aspects of entanglement in the following.

All these aspects of entanglement are quantified by measures that are based on the negativity or the average negativity. The average negativity $\bar{\cal N}$ is the negativity averaged over all possible bipartitions of the system~\cite{calsamiglia_spin_gases}. Non-zero average negativity ensures the presence of some form of entanglement in the system.

Specifically, we study three types of entanglement:
\begin{enumerate}
\item The average negativity of the density matrices with $n$ qubits, averaged over the Poissonian distribution of the number of particles in the system: $\langle\bar{\cal N}(\rho_n)\rangle_{p_\lambda(n)}$.

\item The negativity of reduced $2$-qubit density matrices averaged over a renormalized, truncated Poissonian distribution: $\langle{\cal N}(\rho_{n\rightarrow 2})\rangle_{\tilde{p}_\lambda(n\geq2)}$.

\item The negativity of averaged, reduced $2$-qubit density matrices: ${\cal N}(\bar{\rho})$.
\end{enumerate}

Now, we lay out, what these quantities mean, and which aspect of entanglement they describe.\\

(i) If we ask how much entanglement we find in the system on average we are led to the quantity $\langle\bar{\cal N}(\rho_n)\rangle_{p_\lambda(n)}$, which is the expectation value of the average negativity of density matrices with different $n$, where $p_\lambda(n)$ is the Poissonian probability distribution for the number $n$ of particles in the system.
Observe that if we disallowed the fluctuation of the particle number in the system and introduced the reset mechanism by other means (e.g. measurement and decay to standard state inside the box), the quantity of interest would simply be $\bar{\cal N}(\rho_N)$ for fixed system size $N$.

(ii) When we look at subsystems we are led to slightly different quantities. Let us fix the subsystem size to $2$ qubits. We call the reduced density matrices of originally $n$ qubits $\rho_{n\rightarrow 2}$ where we assume $n\geq 2$. In gas-type systems there can also be zero or one particle in the box (especially if $\eta\propto\lambda$ is small), and the entanglement is simply zero in these cases. Since a ``reduction" of a one or zero-qubit density matrix to a $2$-qubit density matrix makes no sense, we simply exclude these cases and rescale the truncated Poissonian $p_\lambda(n\geq2)$ to the distribution $\tilde{p}_\lambda(n\geq2)$. The quantity $\langle{\cal N}(\rho_{n\rightarrow 2})\rangle_{\tilde{p}_\lambda(n\geq2)}$ therefore tells us how much entanglement a subsystem of two qubits contains on average (for $2$ qubits ${\cal N}=\bar{\cal N}$).

(iii) When we look only at a subsystem of two qubits disregarding the number of particles $n$ in the system, then our best description of the $2$-qubit density matrix is the average density matrix $\bar{\rho}:=\langle\rho_{n\rightarrow 2}\rangle_{\tilde{p}_\lambda(n\geq2)}$ with entanglement ${\cal N}(\bar{\rho})$. \\

When we compare the three kinds of entanglement defined above we see that the conditions that one of these quantities be non-zero are increasingly stringent. To find entanglement in the reduced system is a more stringent condition since tracing out the other particles has the same effect as an additional noise source. Also, to find entanglement in the averaged density matrix $\bar{\rho}$ is a stricter condition since the averaging increases entropy, i.e., tends to make the matrix more mixed. If we keep the particle number fixed, $N\geq 2$, there is just the quantity ${\cal N}(\rho_{N\rightarrow 2})$ to describe the entanglement in the reduced state.

\begin{figure}[ht]
\includegraphics[width=0.7\textwidth]{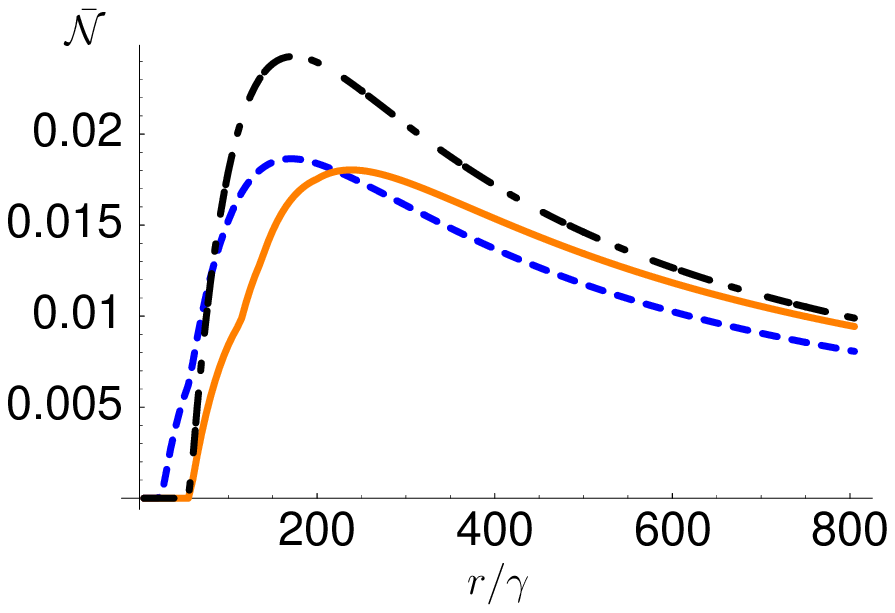}
\caption[]{\label{morequbits}Measures $\langle\bar{\cal N}(\rho_n)\rangle_{p_\lambda(n)}$ (blue, dashed), $\langle{\cal N}(\rho_{n\rightarrow 2})\rangle_{\tilde{p}_\lambda(n\geq2)}$ (orange, solid), and ${\cal N}(\bar{\rho})$ (black, dashed-dotted) as functions of the reset rate $r/\gamma$. The particle number fluctuates between two and five, and the fluctuations are accounted for by a Poissonian or truncated Poissonian weighting with $\lambda=2$ as explained in the text. The interaction strength of the Ising interactions between all particles is $g=20\gamma$, while the strength of the local, free Hamiltonian~(\ref{atomicHamiltonian}) is $\omega=50\gamma$. Kinks in the orange curve stem from the averaging over negativities whith different supports (see (ii) in the text).}
\end{figure}

To compute the average negativity is a hard task. The system size, and hence the number of differential equations we must solve, and also the number of bipartitions scale exponentially. To simplify the computation, we consider a symmetric situation, where all qubits interact pairwise via Ising interactions and are subject to dephasing noise~(\ref{dephasingchannel}). In Figure~\ref{morequbits} we plot the three entanglement measures $\langle\bar{\cal N}(\rho_n)\rangle_{p_\lambda(n)}$ (blue, dashed curve), $\langle{\cal N}(\rho_{n\rightarrow 2})\rangle_{\tilde{p}_\lambda(n\geq2)}$ (orange, solid curve), and ${\cal N}(\bar{\rho})$ (black, dashed-dotted curve) for $\lambda=2$ and for a qubit number that fluctuates between two and five. Since the meaning of these measures is different, one cannot compare the absolute values represented by the curves directly with one exception. The points, where the curves become non-zero, must, from left to right, appear in the order explained in the previous paragraph, i.e., blue first, representing measure (i), orange next, representing (ii), and black last, representing (iii).

\subsection{Strongly coupled systems}\label{stronglycoupledmultipartite}

Eventually, we study equations~(\ref{MEsolidstate}) and~(\ref{reset_strongly_coupled_systems}) in the multipartite case. We obtain similar results as in the case of gas-type systems underlining again how generic the reset mechanism is.

\begin{figure}[ht]
\includegraphics[width=0.7\textwidth]{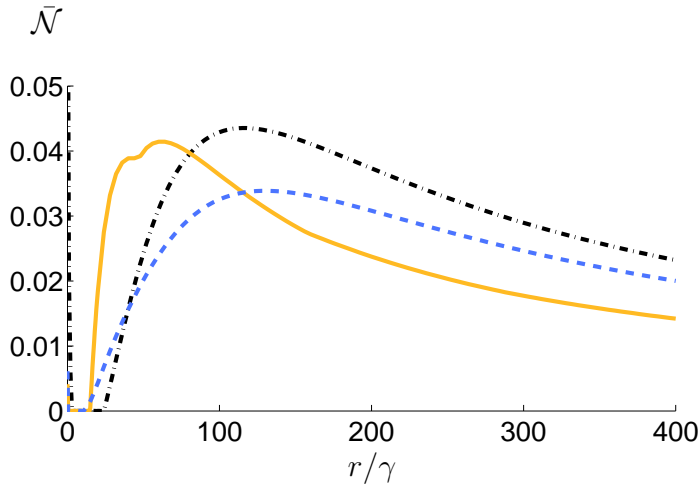}
\caption[]{\label{5qubitsStronglycoupled}Measures $\langle\bar{\cal N}(\rho_n)\rangle_{p_\lambda(n)}$ (blue, dashed), $\langle{\cal N}(\rho_{n\rightarrow 2})\rangle_{\tilde{p}_\lambda(n\geq2)}$ (orange, solid), and ${\cal N}(\bar{\rho})$ (black, dashed-dotted) as functions of the reset rate $r/\gamma$. The particle number fluctuates between two and five, and the fluctuations are accounted for by a Poissonian or truncated Poissonian weighting with $\lambda=2$. The other parameters of the Hamiltonian~(\ref{HisingGradientMagnField}) are $g=15\gamma$ and $b=0.1$. The inverse temperature is $\beta=0.2$. Kinks in the orange curve stem from the averaging over negativities whith different supports (see (ii) in the text).}
\end{figure}

Although a fluctuation of particles may seem less natural in strongly coupled systems as compared to gas-type systems, we will look at the exact same entanglement quantities, so that the corresponding plots are directly related to each other. As Hamiltonian, we choose a sum of pairwise Ising interactions and magnetic fields in $x$ and $z$ direction according to
\begin{equation}\label{HisingGradientMagnField}
H=g\left[\left(\sum_{i>j}\sigma_z^{(i)}\sigma_z^{(j)}\right)+b\sum_{k=1}^N\left(\sigma_x^{(k)}+10^{-5}\frac{k}{N}\sigma_z^{(k)}\right)\right]
\end{equation}
where the small gradient magnetic field in $z$ direction is introduced for technical reasons to lift degeneracies in the Hamiltonian. Figure~\ref{5qubitsStronglycoupled} shows a plot of the same entanglement measures as in the previous subsection. As in the gas-type scenario, the feature that entanglement can be created by a reset mechanism holds also in the multipartite case. While we have shown the genericity of the reset mechanism with respect to the Hamiltonian and noise process already in previous sections, here we demonstrate that the reset mechanism is also generic with respect to system size.

\section{Summary}\label{summary}

We have shown that entanglement can be present in dissipative, open quantum systems far from thermodynamic equilibrium if we assume the existence of an additional mechanism that ``resets" the particles, at a certain rate, into a single-particle, low-entropy state. For a $2$-qubit toy model of a gas-type system, we have analytically solved the master equation consisting of a Hamiltonian part, a noise channel, and the proposed reset mechanism. For special cases we have been able to give closed expressions for the entanglement as a function of the parameters of the master equation. We have extended the analysis to similar models with other interaction Hamiltonians, decoherence models, and imperfect reset mechanisms. We have treated the situation of strongly correlated systems by the same means and we have given conditions under which steady-state entanglement arises in this case. Finally, we have shown that in systems consisting of more qubits, and even in systems with fluctuating particle number, steady-state entanglement can prevail.

Many systems are conceivable for an experimental realization of such dissipative, open quantum systems. For instance, one may consider ions in microtraps that interact via an induced dipole moment~\cite{cirac_zoller}
leading effectively to a continuously operating Ising interaction. Decoherence, dominated by dephasing noise, appears naturally in such systems, and the reset mechanism may, e.g., be achieved by periodically applying a $\pi$-pulse that couples the internal level $|1\rangle$ to a metastable auxiliary level $|a\rangle$ that decays rapidly to $|0\rangle$. The state afterwards is always $|0\rangle$, which can be mapped to $|+\rangle$ by a subsequent Hadamard operation.

For charge manipulated quantum dots, the exchange interaction leads to a continuously operating Heisenberg interaction between neighboring electron spins by lowering the potential barrier~\cite{loss_qc_quantum_dots}. The effect of surrounding nuclear spins may be described by dephasing noise, while the reset mechanism can consist in replacing an electron by a fresh one from the surrounding Fermi sea, prepared in a suitable state (e.g., $|0\rangle$).

Atomic beams interacting via a cavity mode~\cite{raimond_cavity_qed} may also serve as a toy example of such systems far from thermodynamic equilibrium. 

Note that a reset mechanisms could be realized in many physical ways, including a measurement with subsequent preparation of the state, coupling or decay to metastable auxiliary states, as well as replacing a qubit by a fresh one. While these suggestions for implementations aim at demonstrating how a reset mechanism would be realized in experiments, the effect itself is generic and other realizations are conceivable. In particular, one might try to find such a reset mechanism in less controlled, maybe even biomolecular systems consisting of many particles.

\ack
This work was supported by the Austrian Science Foundation (FWF), the European Union (\textsc{olaqui}, \textsc{scala}, \textsc{qics}) and the \"OAW through project \textsc{apart} (W.D.).

\appendix
\section{}\label{appendix}
In this appendix we derive the solution of the master equation~(\ref{gasME}) for the Ising Hamiltonian~(\ref{Isinghamiltonian}) explicitly. In the final formulae we restrict ourselves again to the dephasing channel~(\ref{dephasingchannel}).

Instead of solving the master equation by spectral decomposition of the matrix $\Lambda$ associated with the Liouville operator ${\cal L}$ (see section~\ref{SolutionMEgastype}), we solve the system of linear differential equations step by step. Since not all of the differential equations are coupled, this leads to simpler expressions. The disadvantage is, however, that the set of solutions contains $9$ integration constants that must be determined afterwards for given initial conditions.

We expand the $2$-qubit density matrix $\rho$ in the standard basis $|\mathbf{s}\rangle=|s_1s_2\rangle$ with $s_j\in {0,1}$ and $\sigma_z^{(j)}|\mathbf{s}\rangle=(-1)^{s_j}|\mathbf{s}\rangle$. The expression for the density matrix becomes $\rho=\sum_{s_1^\prime s_2^\prime s_1s_2}C_{s_1^\prime s_2^\prime s_1s_2} |s_1^\prime s_2^\prime\rangle\langle s_1s_2|$. Inserting this expansion into the master equation, and defining the two functions
\begin{eqnarray*}
f_1(x^\prime, x)&:=&\textstyle\frac{B}{2}(1-s)[(1-x^\prime)+(1-x)]\\ &-&\textstyle\frac{B}{2}s[(1-(x^\prime\oplus1))+(1-(x\oplus1))]-\textstyle\frac{2C-B}{4}[1-(-1)^{x^\prime+x}],\\
f_2(x^\prime,x)&:=&Bs(1-x^\prime)(1-x)+B(1-s)(1-(x^\prime\oplus1))(1-(x\oplus1)),
\end{eqnarray*}
we get the following linear system of coupled differential equations
\begin{eqnarray}\label{MEinstandardbasis}
\mbox{d}_t C_{s_1^\prime s_2^\prime s_1s_2}&=&
\{-ig\left[(-1)^{s_1^\prime+s_2^\prime}-(-1)^{s_1+s_2}\right]\nonumber\\
&-&i\omega/2 \left((-1)^{s_1^\prime}-(-1)^{s_1}+(-1)^{s_2^\prime}-(-1)^{s_2}\right)\nonumber\\
&+&f_1(s_1^\prime,s_1)+f_1(s_2^\prime,s_2)-2r\}C_{s_1^\prime s_2^\prime s_1s_2}\nonumber\\
&+&f_2(s_1^\prime,s_1)C_{(s_1^\prime\oplus1) s_2^\prime (s_1\oplus1) s_2}+f_2(s_2^\prime,s_2)C_{s_1^\prime (s_2^\prime\oplus1) s_1 (s_2\oplus1)}\nonumber\\
&+&r/2\{C_{0 s_2^\prime 0 s_2}+C_{1 s_2^\prime 1 s_2}+C_{s_1^\prime 0 s_1 0}+C_{s_1^\prime 1 s_1 1}\}.
\end{eqnarray}
Here, the operation $\oplus$ means addition modulo $2$. Fortunately, these $16$ differential equations are not fully coupled. The coefficients $C_{0000}$, $C_{0101}$, $C_{1010}$, and $C_{1111}$ on the diagonal of the density matrix are coupled only to themselves. Once we have solved these equations, we can treat the diagonal coefficients as known inhomogeneities in the other equations. The off-diagonal coefficients $C_{0001}$ and $C_{1011}$ are coupled among themselves and to the diagonal, so we can solve them next. The same is true for the pair $C_{0010}$ and $C_{0111}$. Finally, the anti-diagonal coefficients $C_{0011}$ and $C_{0110}$ are coupled to $C_{0001}$,$C_{1011}$,$C_{0010}$, and $C_{0111}$, or to their complex conjugates. We solve these as a last step, and all other coefficients are given by the Hermiticity of the density matrix. The solution is now straightforward in principle. However, the expressions for the matrix coefficients are still space-consuming, so we will give them only for the special case of the dephasing channel.\\

For the dephasing channel~(\ref{dephasingchannel}), the structure of the differential equations is still the same and not simplified, we save space only because we have fewer parameters and a symmetric situation. The solution for the diagonal elements then reads:
\begin{eqnarray*}
C_{0000}&=&\frac{1}{4}+\frac{1}{4}D_2e^{-2rt}+\frac{1}{2}D_3e^{-rt}\\
C_{0101}&=&\frac{1}{4}-\frac{1}{4}D_2e^{-2rt}+\frac{1}{2}D_4e^{-rt}\\
C_{1010}&=&\frac{1}{4}-\frac{1}{4}D_2e^{-2rt}-\frac{1}{2}D_4e^{-rt}\\
C_{1111}&=&\frac{1}{4}+\frac{1}{4}D_2e^{-2rt}-\frac{1}{2}D_3e^{-rt}\\
\end{eqnarray*}

The integration constants $D_2$, $D_3$, and $D_4$ accommodate the initial conditions. Since $\mbox{tr}\rho=1$ is a constraint, there is no free constant $D_1$.

The off-diagonal elements are:

\begin{eqnarray*}\fl
C_{0001}&=&\frac{r \left(-i g+r+\gamma +\frac{i \omega }{2}\right)}{4 \left(2 g^2+\left(r+\gamma +\frac{i\omega }{2}\right) (r+2 \gamma +i \omega )\right)}-\frac{(\text{D}_3+\text{D}_4) e^{-r t} r (2 i g-2 \gamma -i \omega )}{4 \left(4 g^2+(2 \gamma +i \omega ) (r+2\gamma +i \omega )\right)}\\
\fl&-&e^{-\frac{1}{2} t \left(3 r+4 \gamma +2 i \omega +\sqrt{r^2-16 g^2}\right)}\\ \fl&\times&\left(e^{\sqrt{r^2-16 g^2} t} \text{OD}_2 \left(\sqrt{r^2-16 g^2}-4 i g\right)+\text{OD}_1 \left(4 i g+\sqrt{r^2-16 g^2}\right)\right)
\end{eqnarray*}

\begin{eqnarray*}
\fl C_{1011}&=&\frac{r\left(i g+r+\gamma +\frac{i \omega }{2}\right) }{4 \left(2 g^2+\left(r+\gamma +\frac{i \omega }{2}\right) (r+2 \gamma +i \omega)\right)}-\frac{(\text{D}_3+\text{D}_4) e^{-r t} (2 i g+2 \gamma +i \omega ) r}{4 \left(4 g^2+(2 \gamma +i \omega ) (r+2 \gamma +i \omega )\right)}\\
\fl&+&e^{-\frac{1}{2} t \left(3 r+4 \gamma +i \left(2 \omega +\sqrt{16 g^2-r^2}\right)\right)} r\left(\text{OD}_1-e^{i \sqrt{16 g^2-r^2} t} \text{OD}_2\right)
\end{eqnarray*}

The coefficients $C_{0010}$, $C_{0111}$ are very similar to $C_{0001}$, $C_{1011}$, except that $\text{D}_4$ must be replaced by $\text{-D4}$,  and the integration constants are $\text{OD3}$, $\text{OD4}$ instead of $\text{OD}_1$, $\text{OD}_2$.
Finally, the elements on the anti-diagonal are

\begin{eqnarray*}
\fl C_{0011}&=&e^{-2 t (r+2 \gamma +i \omega )} \text{AD}_1
+\frac{(2 r+2 \gamma +i \omega ) r^2}{4 \left(4 g^2+2 r^2+(2 \gamma +i \omega )^2+r (6 \gamma +3 i \omega )\right) (r+2 \gamma +i \omega )}\\
\fl&-&\frac{i e^{-r t} g D_3 r^2}{\left(4 g^2+(2 \gamma +i \omega ) (r+2 \gamma +i \omega )\right) (r+4 \gamma +2 i \omega )}+e^{-\frac{1}{2} t \left(3 r+4 \gamma +2 i \omega +i \sqrt{16 g^2-r^2}\right)}\\
\fl&\times&r\Bigg(\frac{\left(4 g+i r+\sqrt{16 g^2-r^2}\right) \left(\text{OD}_1+\text{OD}_3\right)}{i r+4 i \gamma -2 \omega +\sqrt{16 g^2-r^2}}\\
\fl&&\quad-\frac{e^{i\sqrt{16 g^2-r^2} t} \left(-4 g-i r+\sqrt{16 g^2-r^2}\right) \left(\text{OD}_2+\text{OD}_4\right)}{-i r-4 i \gamma +2 \omega +\sqrt{16 g^2-r^2}}\Bigg)
\end{eqnarray*}
and
\begin{eqnarray*}
\fl C_{0011}&=& e^{t (-2 r-4 \gamma )} \text{AD}_2\\
\fl&+&\frac{ r^2\left(8 (r+\gamma ) g^2+(r+2 \gamma ) \left(4 (r+\gamma )^2+\omega ^2\right)\right)}{4 (r+2 \gamma ) \left(\omega ^4+\left(-8 g^2+5 r^2+8 \gamma ^2+12 r\gamma \right) \omega ^2+4 \left(2 g^2+(r+\gamma ) (r+2 \gamma )\right)^2\right)}\\
\fl&+&\frac{i e^{-r t} g \left(i (r+4 \gamma ) \omega  D_3+\left(4 g^2-\omega ^2+2\gamma  (r+2 \gamma )\right) D_4\right) r^2}{(r+4 \gamma ) \left(16 g^4+8 \left(2 \gamma  (r+2 \gamma )-\omega ^2\right) g^2+\left(4 \gamma ^2+\omega ^2\right)\left((r+2 \gamma )^2+\omega ^2\right)\right)}\\
\fl&+&e^{-\frac{1}{2} t \left(3 r+4 \gamma -2 i \omega -i \sqrt{16 g^2-r^2}\right)}r\Bigg(\frac{\left(4 g-i r+\sqrt{16g^2-r^2}\right) \text{OD}_1}{-i r-4 i \gamma +2 \omega +\sqrt{16 g^2-r^2}}\\
\fl&&-\frac{e^{-i \sqrt{16 g^2-r^2} t} \left(-4 g+i r+\sqrt{16 g^2-r^2}\right)\text{OD}_2}{i r+4 i \gamma -2 \omega +\sqrt{16 g^2-r^2}}\Bigg)\\
\fl&+&e^{-\frac{1}{2} t \left(3 r+4 \gamma +2 i \omega +i \sqrt{16 g^2-r^2}\right)}r\Bigg(\frac{\left(4 g+i r+\sqrt{16 g^2-r^2}\right) \text{OD}_3}{i r+4 i \gamma +2 \omega +\sqrt{16 g^2-r^2}}\\
\fl&&-\frac{e^{i \sqrt{16 g^2-r^2} t} \left(-4 g-i r+\sqrt{16 g^2-r^2}\right) \text{OD}_4}{-i r-4 i \gamma -2 \omega +\sqrt{16 g^2-r^2}}\Bigg)
\end{eqnarray*}
All other coefficients follow from the Hermiticity of the density matrix.
The form of all matrix coefficients is similar. First, there are the parts with the integration constants that fall off exponentially with time. The characteristic exponents are the eigenvalues of the homogeneous parts of each linear differential equation or system of equations (multiplied with time). As pointed out in~\ref{gastype_solutionIsing}, these exponents are the spectrum of the total Liouville super-operator defined by $\dot{\rho}={\cal L}\rho$ with values
\begin{eqnarray*}
\fl\{0,-r, -2r, -2(r+2\gamma), -2(r+2\gamma\pm i\omega),\\ \fl-1/2(3r+4\gamma+\sqrt{-16g^2+r^2}\pm2i\omega,-1/2(3r+4\gamma-\sqrt{-16g^2+r^2}\pm2i\omega\}
\end{eqnarray*}
and multiplicities $\{1,2,1,2,1+1,2+2,2+2\}$ respectively. Second, there always is a part independent of $t$ (belonging to the eigenvalue $0$) that represents the value in the steady state.

\section{}\label{appendix2}

In this appendix we present the steady-state solution for the master equation~(\ref{GeneralME}) with Ising Hamiltonian, local noise channels as in equation~(\ref{localnoise}) and reset states $|\chi_j\rangle=|+\rangle$ to~\ref{appendix2}. The solution in form of the matrix coefficients in the computational basis is given by:

\begin{eqnarray*}\label{equilibriumstate}
\fl C_{0000}=\frac{(r+2 B s)^2}{4 (B+r)^2}\\
\fl C_{0101}=C_{1010}=\frac{(r+2 B (1-s)) (r+2 B s)}{4 (B+r)^2}\\
\fl C_{1111}=\frac{(r+2 B (1-s))^2}{4 (B+r)^2}\\
\fl C_{0001}=\frac{r (r+2 B s) (B+C-2 i g+2 r+i \omega )}{4 (B+r) \left(4 g^2+(C+r+i \omega ) (C+2 r+i \omega )+B (C+r+2 i g (2 s-1)+i \omega )\right)}\\
\fl C_{0011}=\Big\{4 (B+r) \left(4 g^2+(C+r+i \omega ) (C+2 r+i \omega )+B (C+r+2 i g (2 s-1)+i
   \omega )\right) \\
\fl\quad\quad\quad\times(C+r+i \omega )\Big\}^{-1}\Big\{r^2 \left(B^2+(C+2 i g+3 r-4 i g s+i \omega ) B+r (C+2 r+i \omega )\right)\Big\}\\
\fl C_{0110}=\Big\{4 (B+r) (C+r) \left(\omega ^4+\left(2 C^2+6 r C-8 g^2+5 r^2\right) \omega ^2\right.\\
\fl\left.+\left(4 g^2+(C+r) (C+2 r)\right)^2+2 B \left((C+2 r) \omega ^2+2 g (2 C+3 r) (2 s-1)\omega \right.\right.\\
\fl\left.\left.+(C+r) \left(4 g^2+(C+r) (C+2 r)\right)\right)+B^2 \left((C+r)^2+(g (4 s-2)+\omega )^2\right)\right)\Big\}^{-1}\\
\fl\times\Big\{r^2 \left((C+r) B^3+\left((C+r) (2 C+5 r)-16 g^2 (s-1) s\right) B^2\right.\\
\fl\left.+\left(C^3+7 r C^2+4 g^2 C+14 r^2 C+8 r^3+(C+r) \omega ^2+12 g^2 r-4 g (C+r) (2 s-1) \omega\right) B\right.\\
\fl\left.+r (C+r) \omega ^2+r (C+2 r) \left(4 g^2+(C+r) (C+2 r)\right)\right)\Big\}.
\end{eqnarray*}
Furthermore, $C_{0010}=C_{0001}$, and $C_{0111}=C_{1011}$ are obtained from $C_{0001}$ by replacing $g\rightarrow-g$, $s\rightarrow(1-s)$ in the numerator. The other coefficients are given by the Hermiticity of the density matrix.
Observe that the dephasing and the depolarizing channels are included as special instances of the parameters $B,C,s$ in this analytic expression. The dephasing channel results from putting $B=0$, $s$ drops out, and renaming $C=2\gamma$. The depolarizing channel is given by $s=1/2$, $B=C$, and renaming $C=4\gamma/3$.

The plot in Figure~\ref{QOMEequState} is based on this solution.


\section*{References}

\end{document}